\documentclass[%
 aip,
 amsmath,amssymb,
 reprint,%
]{revtex4-1}

\usepackage{dcolumn}
\usepackage{bm}

\usepackage[utf8]{inputenc}
\usepackage[T1]{fontenc}
\usepackage{mathptmx}
\usepackage{color}
\usepackage[dvipsnames]{xcolor}
\usepackage{amsmath}
\usepackage{physics}
\usepackage{longtable}
\usepackage[pdftex]{graphicx}
\usepackage{enumerate}
\usepackage[normalem]{ulem}

\begin{document}
\title{Sensitivity and Dimensionality of Atomic Environment Representations \\ used for Machine Learning Interatomic Potentials} 

\author{Berk Onat} 
 \email{b.onat@warwick.ac.uk}
\affiliation{Warwick Centre for Predictive Modelling, School of Engineering, University of Warwick, Coventry CV4 7AL, United Kingdom}

\author{Christoph Ortner}
\affiliation{Warwick Mathematics Institute, University of Warwick, Coventry CV4 7AL, United Kingdom}

\author{James R. Kermode}
 \email{j.r.kermode@warwick.ac.uk}
\affiliation{Warwick Centre for Predictive Modelling, School of Engineering, University of Warwick, Coventry CV4 7AL, United Kingdom}

\date{\today}

\begin{abstract}
Faithfully representing chemical environments is essential for describing materials and molecules with machine learning approaches. 
Here, we present a systematic classification of these representations and 
then investigate: (i) the sensitivity to perturbations and (ii) the effective dimensionality of a variety of atomic environment representations, and over a range of material datasets.
Representations investigated include Atom Centred Symmetry Functions, Chebyshev Polynomial Symmetry Functions (CHSF), Smooth Overlap of Atomic Positions, Many-body Tensor Representation and Atomic Cluster Expansion.
In area (i), we show that none of the atomic environment representations are linearly stable under tangential perturbations, and that for CHSF there are instabilities for particular choices of perturbation, which we show can be removed with a slight redefinition of the representation. In area (ii), we find that most representations can be compressed significantly without loss of precision, and further that selecting optimal subsets of a representation method improves the  accuracy of regression models built for a given dataset.
\end{abstract}
\maketitle

\section{Introduction}\label{intro}

Machine Learning (ML) as a predictive modelling tool has gained much attention
in recent years in fields ranging from biology\cite{BIOINFORMATICS}, and chemistry\cite{CHEMINFORMATICS} to
materials science\cite{Hill2016,Bartok2017}, building on decades of successful applications in image recognition, natural language processing and artificial intelligence (AI) \cite{DeepLearningBook,DeepImageMiningBook}
While machine learning has also been increasingly popular in many 
fields due to its relatively simplicity of application and powerful prediction features, a key driving force of is the increasing 
availability of information through data repositories, 
archives and databases of materials and molecule that include billions 
of structures\cite{Ruddigkeit2012,Hill2016}. In recent years, a large number 
of ML models for materials and molecules have been developed and many novel 
representation methods have been proposed \cite{Behler2011,Bartok2013,Huo2017,Isayev2017,Schutt2018,Zhou2018,Ziletti2018,Novoselov2019}.  
Representations are the feature sets used as input data for data-driven 
machine learning models, which, after training, can be used to predict quantities of interest.

Some of these databases are focused on molecular configurations such as PubChem\cite{PubChem}, 
DrugBank\cite{DrugBank} and ChEMBL\cite{ChEMBL} are widely used in biological, chemical, 
and pharmacological applications driven 
by high-throughput screening for drug design and novel molecule 
discovery. Other databases such as the Cambridge Structural Database (CSD)\cite{CSD} include crystals as well as small molecules. While these databases 
contain mostly molecular structures with their 
chemical or biological properties, recent efforts such as
the Materials Project\cite{Jain2013,MATERIALSPROJECT}, Automatic FLOW for Material Discovery 
Library (AFLOWLIB.org)\cite{Curtarolo2012,Curtarolo2012a}, 
Open Quantum Materials Database (OQMD)\cite{Saal2013,Kirklin2015},
and the NOMAD archive\cite{NOMAD} (which collates 
data from many other databases as well as 
contains data that are not available at them) now provide 
extensive electronic structure 
results  for molecules, bulk materials, surfaces and nanostructures in a range of ordered and disordered phases, with chemical composition ranging from inorganic systems to metals, alloys, and semiconductors. With  recent advancements in 
materials databases, access is available to billions of properties of materials and molecules and millions of high-accurate calculations through online archives. With 
such an extensive and diverse collection of molecular and materials data to hand, we can ask questions such as how to identify of the most informative subset of  data for a particular class of materials, and how best to interpret it for 
prediction through ML models. These questions are relevant for both classification and regression applications: for example, classifying materials as metallic or semiconducting, or predicting the band-gap from structure, respectively. In both cases, the input data is constructed from information ranging from basic physical and chemical properties to precise geometrical information based on atomic
structure and bonding topology. There have been many studies to determine these input information which are named as descriptors, fingerprints or representations \cite{Chen2018,Geiger2013,Collins2018,Muratov2020,Reveil2018,Zhou2018,Behler2016}

Here, we define the prediction problem  as,
\begin{equation}\label{eq1}
{\bf t}(\chi_k) \simeq f(\{M_a(\{{\bf V}_b\})\})
\end{equation}
\noindent where $\mathbf{t}$ is the target property of the material, $\chi_k$ represents the structure  with index $k$ within the database,  $M_a$ is the machine 
learning model with identifier $a$, and ${\bf V}_b$ is the input representation determined with identifier $b$ that is used for optimizing $f$. 
We define {\em representations} as
\begin{equation}\label{eq2}
{\bf V}_b(\chi_k) = g(\{h_i(x_j,\{\alpha_1,\alpha_2,...,\alpha_m\})\}) 
\end{equation}
\noindent where, $x_j$ are the coordinates of atom with index $j$ within structure $k$, $\alpha_m$ 
is a physical or chemical property of the material and chemical environment, $h_i$ is 
the {\em descriptor} mapping from the input space to hyper-dimensional space 
 with each dimension being an input {\em feature} of model $M$ and $g$ is 
the encoding function that combines all of the $h_i$ descriptors to make an overall
{\em representation}. In Equation~\ref{eq1}, the optimization of the parameters of 
$f$ can be performed either for a single model $M$ or for multiple models $M_1, M_2, \ldots$ 
in combination with potentially multiple different representations ${\bf V}_b$.

\begin{figure}
\centering
\resizebox{0.48\textwidth}{!}{\includegraphics{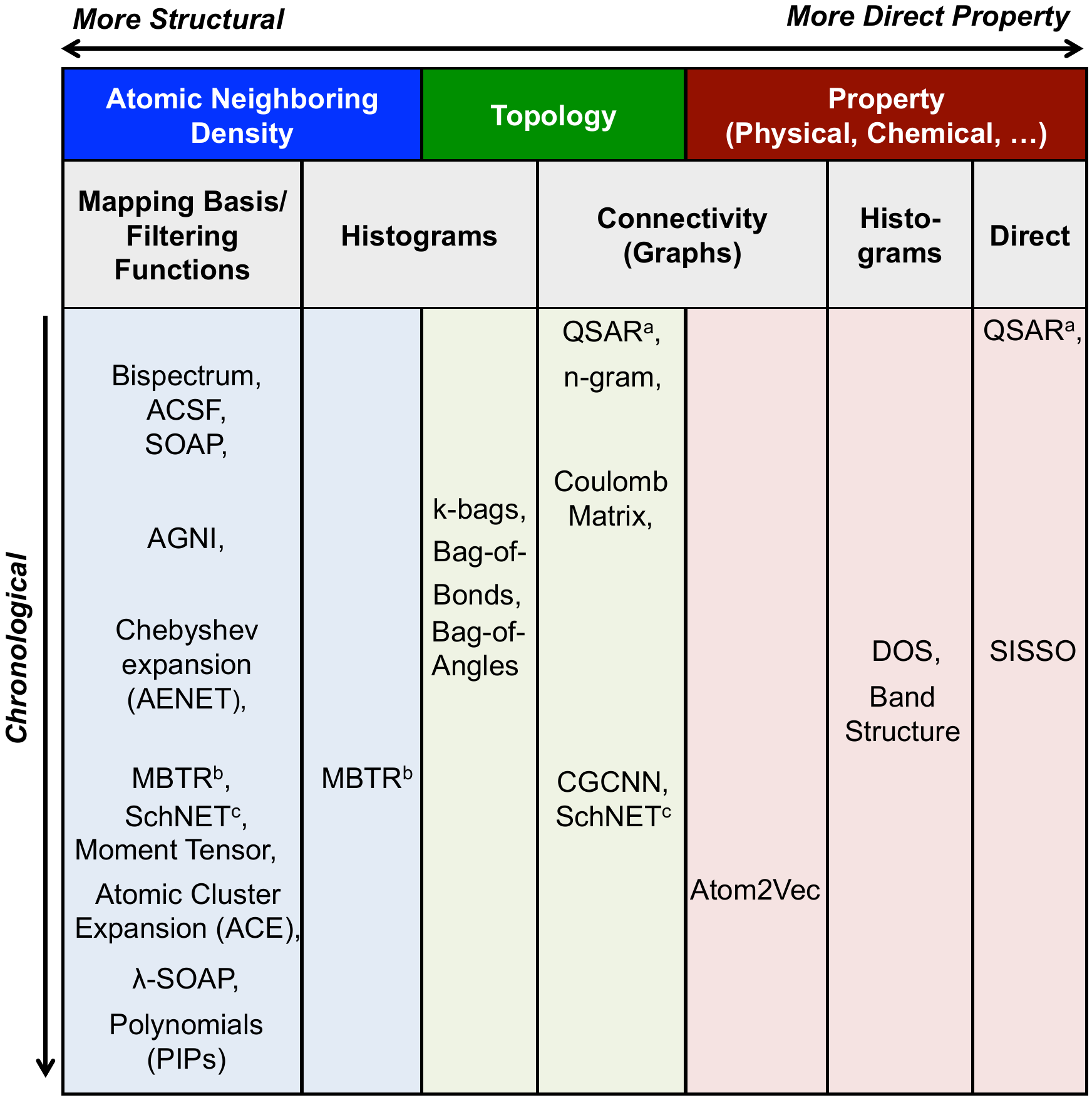}}
\caption{Classification of representations based on their method of construction (horizontal axis) and  when they were first proposed (vertical axis). QSAR and SISSO do not indicate representations but instead indicates the representations that are constructed or selected using these methods (see text). $^{a,b,c}$  representations that are classified with multiple methods: Direct \& Connectivity, Histogram \& Mapping functions, and Connectivity \& Mapping functions.}
\label{fig1}
\end{figure}
 
From this general perspective, many approaches can be combined to form a predictive model, raising the critical open question of how to optimally choose which pieces of information are needed to represent a material through descriptors and how to combine these to form representations.
These can vary from defining sets of 
physical and chemical properties and atomic geometries to specialized hyper-dimensional 
mapping algorithms (see Figure~\ref{fig1}).  These 
representations and their building blocks, the descriptors 
can be classified into three broad 
classes based on their construction: (i) atomic neighborhood density definitions, 
(ii) topology expansions and (iii) property-based selections. 
The representations can be further grouped according to their combination rules,  
mapping basis or filtering functions, histograms, connectivity maps or graphs, and finally direct contributions from material properties. Within each category, there have been various developments of representations and many successful applications, which we review briefly in Section~\ref{classes} below, before specialising on atomic neighbourhood densities in Section~\ref{section:models}. Section~\ref{sec:datasets} lists the data sources used here for evaluating representations, and Section~\ref{sec:analysis} 
describes our analysis methodology. Results and 
discussion follow in Section~\ref{sec:results}.

\section{Representation Classes}\label{classes}

\subsection{Property-based representations}

The idea of classification of molecules based on the relationships between their structure and the resulting activities or properties is an underpinning approach of modern chemistry\cite{QSARbook}. However, 
using theoretical {\em descriptors} to identify the relations was first utilised for quantitative structure-activity/property relationship (QSAR/QSPR) modelling\cite{Lo2018,QSARbook}, starting with the work of Wiener and Platt in 1947, who used indices based on chemical graph theory\cite{QSARbook}. We use the name QSAR refer to both QSAR/QSPR models. 

QSAR models are highly successful predictive modelling approaches 
in chemistry and 
biology in various applications \cite{Lo2018,QSARbook}. These models 
are determined using sets of descriptors that are defined through 
a selection of physical, chemical properties of materials and/or 
structural descriptors of molecules
such as atoms and their molecular bonds from chemical {\em topology} 
information to graphs where the fractions of 
the bonding information is simplified to {\em indices}.\cite{QSARbook,Lo2018,Danishuddin2016}

The success of QSAR mostly depends on the choice of descriptor sets, and
in particular, determining the optimum number of descriptors in such sets \cite{Lo2018}.
For example, compressed-sensing approaches have been developed 
to systematic select the optimal set of the descriptors \cite{Shahlaei2013,Danishuddin2016}.

Direct property-based approaches also employ
reduction techniques to identify optimal models $\{M_a\}$ in Equation~\ref{eq1}, by choosing mathematical operators for $g$ that combine descriptors from 
a given set of $\{h_i\}$ with 
$x_j$, $\{\alpha_m\}$ or both. This 
optimization scheme, 
referred to as {\it sure independence screening and 
sparsifying operator} (SISSO), has been successfully applied as a classification 
technique to group materials (e.g. metals vs. non-metals) and as a regression 
method to determine a target property of interest.\cite{Ouyang2018}

The connectivity information of each species in 
the chemical stoichiometry has also been introduced recently with  
{\em atom2vec}\cite{Zhou2018} method as a representation where a
single atom-to-environment connectivity matrix can be constructed.
This simple approximation enables the approach to be used to screen 
billions of materials and make predictions such as identifying 
possible candidates for Li-ion battery elements \cite{Cubuk2019} 
based only on chemical composition and stoichiometry. 

Recent efforts also address the direct usage of electronic structure 
data to define descriptors.
In these methods, the essence of the electronic structure information 
is extracted through histograms over density-of-states 
(DOS) \cite{Yeo2019} or over the 
selected band structures \cite{BANDS} into the 
descriptor vectors.

\subsection{Topology-based representations}

In this group of descriptors, topological information is derived from the full 
connectivity graph of 
the atomic environment instead of simplified {\it indices}. 
Examples of this class of methods include
the Coulomb and Sine Matrices\cite{Himanen2019}, {\em n}-gram 
graphs\cite{NGRAM,Sutton2019}, and Graph based Neural Network (NN) models such as 
DTNN\cite{Schutt2017}, SchNET\cite{Schutt2018}, CGCNN\cite{CGCNN}. 
While Coulomb and Sine matrices use 
atom-to-atom connection information directly and are thus based on the inferred chemical bonding between atoms, the input representations of Graph Neural Network models 
are defined through neighborhood analysis 
of atomic connectivity graphs and can be constructed from the full $n$-body topology.

Another approach to use topology information such as bonds, angles and higher-order many-body contributions is to group them into sets of building 
blocks and apply histograms on these blocks over the input datasets. Several 
successful implementations of these approach are Bag-of-Bonds (BoB)\cite{Hansen2015} and its 
analogy to higher order contributions for example, Bag-of-Angles and k-bags.

Topology information provides atomic connectivity data and is very useful for 
structural similarity comparison, since this information is automatically 
invariant under changes in the atomic positions. Topology is a 
blueprint of chemical bonding definitions for materials or   
molecules. However, it is not a physical observable and therefore 
it cannot be constructed uniquely.
For novel material discovery one needs to use 
electronic structure calculations to generate new data comprising the 
properties and construction of structural representations of uncharted material compositions. However, such database construction for new structures is extremely expensive.

\subsection{Atomic Neigbourhood Density representations}

Modelling materials and molecules at the atomic scale is essential to identify 
novel materials for applications of interest or to determine 
candidate molecules with specific function in a desired medium. When 
accuracy and reliability of concern, preference is usually given to {\em ab initio} 
calculations. However, these first-principles
calculations are inaccessible for systems requiring long time-scales and/or large 
numbers of atoms. To significantly reduce the computational cost, interatomic 
potentials or force fields are employed to define the interactions between 
atoms using parametrized forms of functions in place of the electronic 
interactions. Although these approaches allow larger numbers of 
atoms to be included in simulations and provide access to much longer time-scales, their prediction 
accuracies are typically limited as a result of 
simplified models describing interatomic interactions.  

A prominent recent trend has been to address this limitation with {\em machine learning interatomic potentials} (MLIPs), i.e. by replacing fixed functional forms for the atomic interactions with data-driven machine learning models that map from atomic positions to potential energy surface, via descriptors based on the local neighbourhood of each atom. This approach based on  atomic-densities extraction has been applied in pioneering models using symmetry functions \cite{Behler2011} 
and more recently the so-called smooth-overlap of atomic positions (SOAP)\cite{Bartok2010,Bartok2013}. Symmetry functions  
are widely used in neural network interatomic potentials 
(NNP)\cite{Behler2011,Behler2016,Artrith2011,Artrith2016} successfully for many molecules and materials 
including water\cite{}, crystals with various 
phases\cite{Cubuk2017}, amorphous solids in various 
concentrations\cite{Sosso2012,Onat2018}, while SOAP is typically incorporated in Gaussian 
Approximation Potentials (GAP)\cite{Bartok2010,Bartok2013}. GAP has been applied to
many molecules and materials.\cite{Dragoni2018,Deringer2019,Bernstein2019,Bartok2017} 

In this final class of descriptors, there have 
been many developments in recent years. In addition to atom-centered symmetry 
functions (ACSF) and SOAP descriptors, different basis expansions such as Bispectrum 
and Chebyshev polynomials have also been utilized in SNAP potentials\cite{Thompson2015,Wood2018} 
and NNPs\cite{Artrith2017}, respectively.  Further advances are also 
provided in ML models based on atomic and many-body expansions with tensor 
representations such as Many-Body Tensor Representation (MBTR)\cite{Huo2017}. 

An alternative approach is to employ linear regression using a symmetric polynomial basis. This approach was pioneered by Bowman and Braams~\cite{BowmanBraams}, while more recent symmetric bases are the Moment Tensor 
potentials (MTP)~\cite{Novoselov2019} and the Atomic Cluster Expansion (ACE)~\cite{ACE}. The MTP and ACE basis are also based on density projections and therefore closely related to the descriptors described above. In particular, ACE can also be seen as a direct generalisation of SOAP and SNAP providing features of arbitrary correlation order.

Recent works have demonstrated that the atomic-density representations can be unified in a common 
mathematical framework\cite{Willatt2019}, and that the descriptors can be decorated with additional properties to extend their capabilities, such as in $\lambda$-SOAP representations for learning from tensor properties\cite{Willatt2019}. 

For all these ML models, different combinations of descriptors and representations 
have been used in many studies and for a wide range of materials. \cite{Deringer2018,Bernstein2019,Zuo2020}
As these descriptors and representations are 
general structural identifiers, they can also be utilized 
within different regression models other than the ML models 
for which they were primarily developed. A recent work also shows the  
comparison of different regressors with these 
representations.\cite{Sutton2019}
While different approaches have been utilized in these studies, comparative assessments 
of the representation approaches have thus far been limited. Studies for the property and topology based 
descriptors on the QSAR feature sets and the performance of graph based NN models show 
that selection of optimal descriptor sets are very important to ensure high predictive accuracy of the ML models\cite{Zuo2020}.
A number of studies of the atomic neighborhood density approaches provide analyses primarily based on trained models' precision on available datasets \cite{Zuo2020}
but so far there is no performance assessment on descriptors and representations for 
materials and molecules focused solely structure to hyper-dimensional encoding.

Following the works by Huo\cite{Huo2017} and J\"ager\cite{Jager2018}, we
identify the essential properties of descriptors and representations
for encoding materials and molecules as follows: 

\begin{enumerate} [(i)]
\item {\it Invariance}: descriptors should be 
  invariant under symmetry operations: permutation of atoms and translation and rotation of 
  structure.
\item {\it Sensitivity (local stability)}: small changes in 
    the atomic positions should result in proportional changes in the descriptor, and vice versa.
\item {\it Global Uniqueness/Faithfulness}: the mapping of the descriptor should be unique for a
given input atomic environment (i.e. the mapping is injective).
\item {\it Dimensionality}: relatedly, the dimension of the spanned hyper-dimensional space of the descriptor should be sufficient to ensure uniqueness, but not larger.
\item {\it Differentiability}: having continuous functions that are differentiable. 
\item {\it Interpretability}: features of the encoding can be mapped directly to 
structural or material properties for easy interpretation of results. 
\item {\it Scalability}: ideally, descriptors should be easily generalized to any 
system or structure with a preference to have no limitations on number of elements, atoms, or properties.
\item {\it Complexity}: to have a low computational cost so the method can be fast 
enough to scale to the required size of the simulations and to be used in high-throughput screening of big-data.
\item {\it Discrete Mapping}: always map to the same hyper-dimensional space with 
constant size feature sets, regardless of the input atomic environment.
\end{enumerate}

In this article, we concentrate our efforts on analysing sensitivity (i.e. local stability) and 
compressibility (i.e. dimensionality) of the selected set of descriptors and representations, and do not address uniqueness/faithfullness; for an interesting recent investigation of this important issue see Ref. \onlinecite{Pozdnyakov2020}.
{\it Scalability}, {\it Complexity}, 
and {\it Discrete Mapping} are also not addressed here
as these are mainly related to the implementation and 
cost of methods in ML models such as the 
evaluation costs of MLIPs or scalability limits of 
ML models, which are outside the scope of this work.
On the other hand, the hunt for 
the {\it interpretability} of ML models  
considering the construction of the descriptors is
not a new concept and has been studied in the context of
QSAR models\cite{Mikulskis2019} and many-body descriptors\cite{Pronobis2018}. As also discussed in 
property-based representations, {\it interpretability} 
is automatically inherited by models that are built directly 
from properties. 
As the concept 
depends on the application of the ML models, 
we leave addressing this for future studies.
While there have been many related efforts on the analysis of property and topology based 
descriptors \cite{Shahlaei2013,Danishuddin2016,Ouyang2018}, we choose to focus here on the atomic neighborhood 
density based descriptors along with representative descriptors from three groups of 
mapping basis functions, tensor representations, and polynomial representations as 
follows: ACSF, Chebyshev polynomials in SF (CHSF), SOAP, SOAPlite, MBTR, and ACE\cite{ACE}. We provide an 
analysis of the sensitivity and local stability of the descriptors as well testing their 
invariance under symmetry operations. We further assess the descriptors information 
packing ability using CUR matrix decomposition, farthest point search (FPS) and 
principal component analysis (PCA) dimension reduction techniques.  We also provide analyses using linear and kernel ridge regression methods on the selected features from CUR 
to showcase the effect of the dimensionality of representations
on ML models.

\section{Atomic Neighborhood Density Representations}\label{section:models}
 
The descriptors sharing neighborhood density extraction can be unified in a general 
atomic expansion following the works in Ref.~\onlinecite{ACE} and \onlinecite{Willatt2019} and 
can be defined by,
\begin{equation}\label{eq3}
V_b = \{b_{i,n}, b_{i+1,n}, ...\} , i=1,..,N
\end{equation}
where
\begin{equation}\label{eq4}
b_{i,n} = \sum_j R_{n}(x_i,x_j).
\end{equation}
Here, $x_i$ is the position of atom $i$ and $b_{i,n}$ can be expressed with radial and angular basis functions or as a polynomial 
expansion. 
In the following, we will give a brief summary of selected atomic neighborhood density 
representations based on the above definition. These are Atom Centered Symmetry 
Functions (ACSF), Chebyshev polynomial representations,
Smooth Overlap of Atomic Positions (SOAP), 
Many-Body Tensor Representation (MBTR), and 
Atomic Cluster Expansion (ACE).
 
\subsection{Atom Centered Symmetry Functions (ACSF)}

The descriptors of atom centered symmetry functions were introduced 
by Behler and Parrinello with their atomic neural 
network potentials (NNP) \cite{Behler2011,Artrith2013} and further
used in a wide variety of applications 
\cite{Behler2011,Artrith2013,Artrith2011,Cubuk2017,Onat2018,Jager2018}. The 
method extracts atomic environment information for each atom in the configuration using 
radial contributions given by 
\begin{equation}\label{eq5}
b^r_i = \sum_{j\neq i}^{N_j} g^r(R_{ij})
\end{equation}
and angular contributions of the form
\begin{equation}\label{eq6}
b^a_i = \sum_{j\neq i}^{N_j}\sum_{k\neq j\neq i}^{N_k} g^a(A_{ijk})
\end{equation}
where the distances $R_{ij}=|\mathbf{R}_j-\mathbf{R}_i|$ and angle dependent contributions are defined through their cosines via $A_{ijk} = \cos\theta_{ijk}$ and $\cos\theta_{ijk}=\mathbf{R}_{ij}\cdot \mathbf{R}_{jk}/(|\mathbf{R}_{ij}| |\mathbf{R}_{ik}|)$
and are invariant under symmetry operations of translations and rotations, 
hence the name {\em symmetry functions}. While many symmetry functions 
have been proposed \cite{Behler2011}, two choices of functions commonly used for $b^r$ and 
$b^a$ in many applications\cite{Behler2011,Artrith2014,Artrith2016,Artrith2017,Onat2018,Imbalzano2018}
take the radial function to be centered on atom $i$ to be defined as
\begin{equation}\label{eq7}
b^r_i=G^2_i = \sum_{j\neq i} e^{-\eta(R_{ij}-R_{s})^2}\cdot f_c(R_{ij})
\end{equation}
and the angular function centred on atom $i$ as
\begin{equation}\label{eq8}
\begin{split}
b^a_i= G^4_i =& 2^{1-\zeta}\sum_{j\neq i}\sum_{k\neq i,j} (1+\lambda A_{ijk})^{\zeta} \\
&\cdot e^{-\eta(R_{ij}^2+R_{ik}^2+R_{jk}^2)} \\
&\cdot f_c(R_{ij})\cdot f_c(R_{ik})\cdot f_c(R_{jk})
\end{split}
\end{equation}
where $f_c$ is a cutoff function given by
\begin{equation}\label{eq9}
f_c(R_{ij}) = 
    \begin{cases}
      0.5\left[\cos\left({{\pi R_{ij}}\over{R_c}}\right)+1\right], & \text{for} \quad R_{ij}\leq R_c, \\
      0, & \text{for} \quad R_{ij}>R_c
    \end{cases}
\end{equation}
and $R_c$ is the cutoff distance.

In this work, we used $G^2$ and $G^4$ with two parameter sets: one is the traditional 
parameter set taken from Ref.~\onlinecite{Behler2011,Artrith2011,Artrith2013,Artrith2016} that is used in many 
ACSF-based NN potential models, and the second one is extracted from the automatic ACSF parameter 
generation proposed in Ref.~\onlinecite{Imbalzano2018}. For the rest of this work, we label
the representation of the original parameter set as ASCF and the 
newer systematic extended parameter set as ACSF-X, for which we take
the same parameters as Ref.~\onlinecite{Imbalzano2018}.

\subsection{Chebyshev Polynomial Representation within Symmetry Functions (CHSF)}

By introducing radial and angular basis that are invariant 
under translation and rotational symmetries, one can define 
different functions that provide invariance under symmetry operations. Another 
example that exploits this idea defines radial and angular functions with 
\begin{equation}\label{eq10}
b^r_i(R_{ij})=\sum_{\alpha=1}^{N_\alpha}c_{\alpha}^{(2)}\phi_{\alpha}(R_{ij}),
\end{equation}
\begin{equation}\label{eq11}
b^a_i(A_{ijk})=\sum_{\alpha=1}^{N_\alpha}c_{\alpha}^{(3)}\phi_{\alpha}(A_{ijk})
\end{equation}
\noindent where $N_\alpha$ is the expansion order and the basis functions $\phi_{\alpha}$ and their duals
$\bar{\phi}_{\alpha}$ are defined in terms of the Chebyshev polynomials $T_{\alpha}$ via

\begin{equation}\label{eq12}
\phi_{\alpha}(x) = {{k}\over{2\pi\sqrt{{x\over{T_c}}-{x^2\over{T_c^2}}}}}
T_{\alpha}\left({{2x}\over{T_c}}-1\right).
\end{equation}
In Equations \ref{eq10} to \ref{eq12}, $\alpha$ is the order of the polynomials, 
$k=1$ except for $\alpha=0$ where $k=2$, and
$(x,T_c)$ are taken to be $(r,r_c)$ for
radial functions and $(A_{ijk},\pi)$ for angular 
functions, respectively.\cite{Artrith2017}
The atomic descriptors $V_b$ are then defined using the set of 
coefficients $c^{(2)}_{\alpha}$ and $c^{(3)}_{\alpha}$ corresponding to $b^r_i$ and 
$b^a_i$ with
\begin{equation}\label{eq13}
c^{(2)}_{\alpha}=\sum_{j\neq i}\phi_{\alpha}(R_{ij})f_c(R_{ij})w_{t_j}
\end{equation}
\begin{equation}\label{eq14}
c^{(3)}_{\alpha}=\sum_{k\neq j\neq i}\phi_{\alpha}(A_{ijk})f_c(R_{ij})f_c(R_{ik})w_{t_j}w_{t_k}
\end{equation}
where $w_t$ is the weight for species $t$. For single-species 
configurations $w_t = 1 $, while for multi-species configurations, both structural and 
compositional parts contribute to the final descriptor.

A practical advantage of using polynomial expansion for radial and 
angular functions is the reduced number of input parameters as the only 
parameter for the expansion is the expansion order $N_{\alpha}$ of the 
Chebyshev polynomials. In this work, we select $N_{\alpha}=9$ 
for both radial and angular polynomial functions.

In this representation, the radial and angular contributions $b^r$ and $b^a$ are separate functions of distances and angles, respectively.
$b^r$ is defined by Equation~\ref{eq10} and provides a {\it histogram of distances} present in the atomic environment.
However, angular contributions $b^a$ are significantly different.
While Chebyshev polynomial variant defines $b^a$ to be a {\it histogram of angles} using only $A_{ijk}$ in Equation~\ref{eq11}, ACSF combines both distances and angles in Equation~\ref{eq8} and thus defines {\it histogram of triangles}.

\subsection{Smooth Overlap of Atomic Positions (SOAP)}

Descriptors can be constructed for extracting neighboring 
atomic environments using the smooth overlap of atomic positions 
(SOAP) approach \cite{Bartok2013}. In this method, atomic densities centered at atom positions are defined by a sum of atom-centered Gaussians with the overall atomic density of a structure $\chi$ is given by
\begin{equation}\label{eq15}
\hat{\rho}_{\chi}(r)=\sum_{i\in \chi}e^{-{{1}\over{2\sigma^2}}|r-R_{i}|^2}
\end{equation}
and one can build SOAP kernel $K(\chi,\chi^{\prime})$ with
\begin{equation}\label{eq16}
K(\chi,\chi^{\prime})=\int d\hat{R}\left|\int\hat{\rho}_{\chi}(r)\hat{\rho}_{\chi^{\prime}}(\hat{R}r)dr\right|^\zeta
\end{equation}
where the exponent $\zeta>1$ and the integral is calculated 
over all possible rotations $\hat{R}$ of the overlapping densities 
of $\chi$ and $\chi^{\prime}$ environments.  In practice 
as is elegantly shown in Ref.~\onlinecite{Bartok2013}, an 
equivalent kernel can be rewritten in the form of 
$K(\chi,\chi^{\prime})=\mathbf{\hat{p}}(\chi)\cdot \mathbf{\hat{p}}(\chi^{\prime})$ by selecting 
a set of orthonormal radial basis functions $g_n(r)$ and angular basis functions with the spherical 
harmonic functions $Y_{lm}(\theta, \phi)$ to expand the atom centered density at atom $i$ with
\begin{equation}\label{eq17}
b^a_i=\rho_{\chi}^{i}(r,\theta,\phi)=\sum_{nlm}c^{i}_{nlm}g_n(r)Y_{lm}(\theta,\phi)
\end{equation}
and using the power spectrum of the expansion coefficients $c^{i}_{nlm}$, given by
\begin{equation}\label{eq18}
\mathbf{\hat{p}}(\chi)=p_{nn^{\prime}l}^{i,j}(\chi)=\pi \sqrt{{8}\over{2l+1}}\sum_{m}(c_{nlm}^i)^*c_{n^{\prime}lm}^j.
\end{equation}
where $n$ and $n^{\prime}$ are indices for the radial basis functions and 
$l$, $m$ are the angular momentum numbers for the spherical harmonics.
In SOAP as it is defined in Ref.~\citenum{Bartok2013}, the radial basis functions are given by
\begin{equation}\label{eq19}
g_n(r)=\sum_{n^{\prime}=1}^{n_{max}}w_{nn^{\prime}}\phi_{n^{\prime}}(r)
\end{equation}
\begin{equation}\label{eq20}
\phi_{n^{\prime}}(r)=(r-R_c)^{n^{\prime}+2}
\end{equation}
in terms of polynomials.  
The representation of atomic environment $\chi$ is then defined by $\mathbf{\hat{p}}(\chi)$, where $p^i(\chi)$ can be identified as  atomic descriptors for atom $i$.
The SOAP descriptors and representation are specified by the expansion orders
$n_{max}$ for the radial basis and $l_{max}$ for the angular basis.
In this work for compatibility with SOAPlite and with the polynomial expansion 
order of Chebyshev polynomials in SF, we select $n_{max}=9$ and $l_{max}=9$.

\subsection{Modified Basis Expansion for SOAP (SOAPlite)}

Introducing a different radial basis function and treatment of 
spherical harmonics basis, a modified version of SOAP referred to as 
SOAPlite has been proposed recently\cite{Jager2018}. In this version of SOAP, radial basis functions 
are replaced by
\begin{equation}\label{eq21}
g_{nl}(r)=\sum_{n^{\prime}=1}^{n_{max}}w_{nn^{\prime}l}\phi_{n^{\prime}l}(r)
\end{equation}
\begin{equation}\label{eq22}
\phi_{n^{\prime}l}(r)=r^{l}e^{-\alpha_{n^{\prime}l}r^2}
\end{equation}
where $\alpha_{nl}$ are decay parameters of non-orthonormal 
functions $\phi_{nl}(r)$ that determines the decay of $\phi_{nl}$ to $10^{-3}$ 
at cutoff radius specified by $(R_c-1)/n_{max}$ steps between 
1\,\AA{} and $R_c$. The method also selects the real (tesseral) spherical 
harmonics for the angular basis as described in Ref.~\citenum{Jager2018}. 
For fair comparison, we select $n_{max}=9$ and $l_{max}=9$ for SOAPlite
with all other parameters taken as the defaults implemented in the 
GAP\cite{Bartok2010,Bartok2013} and QUIP\cite{QUIPcode,Csanyi2007-py} codes.

\subsection{Many-Body Tensor Representation (MBTR)}

Many-body tensor representation (MBTR)\cite{Huo2017} constructs representations 
of structures by defining contributions from $k$ atoms in $k$-body terms with 
$g_k$ geometry functions. In MBTR, these contributions from $k$ atoms 
are smoothed with probability distribution function $D$ and the  
resulting contributions to the representation are given by
\begin{equation}\label{eq23}
f_k(x,z)=\sum_{i=1}^{N_a}w_k(i)D(x,g_k(i))\prod_{j=1}^k C_{z_j,Z_{i_j}}
\end{equation}
where $Z$ are atomic numbers, $C_z$ is an element-element correlation matrix 
consisting of Kronecker $\delta$ values, $w_k$ are weighting functions 
and $g_k$ are scalars for $k$ atoms while $i$ and $j$ are neighbouring atoms in $i=(i_1,\cdots,i_k)$. 
Common selections for the functions $g_k$ are atomic number $g_1(Z_i)=Z_i$, inverse distances of 
$i$-$j$ pairs with $g_2(i,j)=1/|R_i-R_j|$, and angles 
with $g_3(i,j,k)=\angle (R_i-R_j,R_i-R_k)$.
In this work, we select $g^r=g_2$ and $g^a=g_3$ for geometry functions with exponential decay function 
\begin{equation}\label{eq24}
w_k=e^{-\beta|R_i-R_j|}
\end{equation}
where $\beta$ is taken to be 4.0 for 
$k$=2 and 3.0 for $k$=3. We select Gaussian distribution for $D$ and 
the continuous broadened results are discretized to 
$N_x = (x_{\mathrm{max}} - x_{\mathrm{min}})/ \Delta x$ values using $\Delta x$ steps where $N_x$ is
100 and $x_{\mathrm{min}}$, $x_{\mathrm{max}}$ with
intervals of $[0.1,1.1]$ and $[-0.15,\pi+0.1\pi]$ for distances and angles, respectively. MBTR is the only global representation studied in this work, since it does not include any atomic descriptors but instead
provides a representation for a given structure overall.

\subsection{Atomic Cluster Expansion (ACE)}
The ACE~\cite{ACE} method constructs a complete basis of invariant polynomials. Each basis function may be interpreted as an invariant feature, which can then be collected into a descriptor map. Similarly to SOAP, the ACE starts with a density projection,
\begin{equation}\label{eq25}
    b^a(r,\theta,\phi) = C_{nlm} = \sum_{j} g_n(r_j) Y_l^m(\hat{\bm r}_j),
\end{equation}
where $g_n$ is a radial basis. The atomic positions are not smeared as in SOAP, SOAPlite and MBTR. Isometry invariant features are then obtained by integrating the $N$-correlations over the symmetry group: for ${\bm n} = (n_\alpha)_{\alpha = 1}^N, {\bm l} = (l_\alpha)_{\alpha = 1}^N, {\bm m} = (m_\alpha)_{\alpha = 1}^N$ we obtain
\begin{equation}\label{eq26}
    B_{{\bm n \bm l \bm m}} := \int_{{\rm O}(3)} \prod_{\alpha = 1}^N C_{n_\alpha l_\alpha m_\alpha}.
\end{equation}
Finally, one selects a linearly independent subset of the $B_{{\bm n \bm l \bm m}}$ basis functions. 
A detailed description of this construction is provided in Ref \citenum{SHIPs}.

Aside from the lack of smearing and the choice of radial basis the 2-correlation functions are equivalent to SOAP, while the 3-correlation functions are equivalent to SNAP. Since the ACE construction readily applies to higher order correlations we will use up to 5-body correlations in order to test the effect of introducing significantly higher correlations into the descriptor. To control the size of the feature set we use an {\it a priori} chosen sparse selection as described in Ref~\citenum{SHIPs}.

To complete the specification of the ACE descriptors we must define the radial basis: here we choose 
\begin{equation}\label{eq27}
    g_n(r) = P_n(-r^{-2}),
\end{equation}
where $(P_n)$ is a basis of orthogonal polynomials such that $g_n(R_c) = g_n'(R_c) = 0$. 
In this work, the ACE method is used as implemented in the ACE.jl package in Ref \citenum{SHIPscode}. 

In common with ACSF, SOAP and SOAPlite, ACE provides a histogram of triangles by combining radial basis and 
spherical harmonics in Equations~\ref{eq16} and \ref{eq24}. In all representations, we select a cutoff distance of $R_c$=6.5\,\AA{}.

\subsection{Modified Chebyshev Polynomial Symmetry Functions (CHSF-mix)}
Noting the {\it histogram of triangles} provided by ACSF, SOAP, SOAPlite 
and ACE through Equations~\ref{eq8}, \ref{eq17} and \ref{eq25}, 
we examine the contributions from the angular terms with $A_{ijk}$ and 
radial basis of CHSF. Combining radial $b^r$ and angular $b^a$ basis 
expansion of Chebyshev polynomials, we introduced a new $b^a$ that 
combines both with 

\begin{equation}\label{eq28}
b^a_i(R_{ij},A_{ijk})=\sum_{n=1}^{N_n}\sum_{n'=1}^{N_{n}}\sum_{l=1}^{N_l}c_{nn'l}^{(3)}\phi_{n}(R_{ij})\phi_{n'}(R_{ij})\phi_{l}(A_{ijk})
\end{equation}
where $c^{(3)}_{nn'l}$ is defined as
\begin{equation}\label{eq29}
c^{(3)}_{nn'l}=\sum_{k\neq j\neq i}\phi_{n}(R_{ij})\phi_{n'}(R_{ij})\phi_{l}(A_{ijk})f_c(R_{ij})f_c(R_{ik})w_{t_j}w_{t_k}
\end{equation}
and $\alpha$ index is substituted with $l$ for angular basis and $n$, $n'$ for radial basis sets with the choice of $(x,T_c)$ that are selected in $\phi_{\alpha}(R_{ij})$ and $\phi_{\alpha}(A_{ijk})$ as in 
Equations~\ref{eq13} and \ref{eq14}, respectively.

We analyse the benefit of these novel modifications over CHSF-mix
in Section~\ref{sec:Sensitivity_Results} using $N_n=9$ and $N_l=9$.

\subsection{Descriptor implementations}\label{section:codes}

All our analyses are carried out using our DescriptorZoo 
code \cite{descriptorzoo} (github.com/DescriptorZoo)
that includes implementations of the 
CUR, FPS and PCA analyses and uses AMP\cite{AMPcode}, Dscribe\cite{Himanen2019,Dscribe}, 
qmml-pack\cite{MBTRcode}, QUIP\cite{QUIPcode,Csanyi2007-py} and GAP\cite{Bartok2010,Bartok2013} with its 
Python interface quippy, \ae py\cite{AEPYcode} a wrapper code 
for \AE NET\cite{Artrith2016,Artrith2017,AENETcode} Fortran code of NN ML model based on 
ACSF and Chebyshev polynomial descriptors (CHSF), CHSF.jl for both CHSF and CHSF-mix\cite{CHSF} and ACE.jl\cite{SHIPs} code for ACE representation (labelled ACE in results below). 

\section{Datasets of Materials and Molecules for Analysis}\label{sec:datasets}

We used a wide range of materials and molecules databases to provide datasets to test the various
representation methods.  For diversity, we selected a range of materials and molecular 
systems: Si for single species tests, TiO$_{2}$ for metal-oxides, AlNiCu for metals and 
metal alloys, and molecular configurations containing the elements C, H, O, and N.

{\it Si dataset}: This dataset was constructed using the available GAP Si potential database from 
Ref. \onlinecite{Bartok2018} plus Si molecular dynamics (MD) database from Ref. \onlinecite{Cubuk2017}. 
While the overall dataset includes various crystalline phases of Si, it also includes 
MD data. This dataset includes 3,583 structures with 242,139 atomic environments.

{\it TiO$_2$ dataset}: We used a TiO$_2$  dataset that was designed to build atom neural network
potentials (ANN) by Artrith {\em et al.} \cite{Artrith2016,Artrith2017} using the \AE{}NET package. This dataset includes 
various crystalline phases of TiO$_2$ as well as MD data that is extracted from {\em ab inito} calculations. The 
dataset includes 7,815 structures with 165,229 atomic environments in the stochiometric ratio of 66\% O to 34\% Ti.

{\it AlNiCu dataset}: This dataset is formed from two parts: single species datasets for Al, Ni, and Cu 
from the NOMAD Encyclopaedia and multi-species datasets that include Al, Ni and Cu from NOMAD Archive.
All single-specie data was fetched from the NOMAD Encyclopaedia, after removing duplicate 
records with degenerate atomic environments (e.g. equivalent structures from different 
{\em ab initio} calculations uploaded to NOMAD). For the multi-species data, we used only the last configuration steps 
for each NOMAD Archive record, since these records include all intermediate calculation cycles, with the 
last configuration entry typically corresponding to a fully relaxed configuration. In our dataset, 
the NOMAD unique reference access IDs are retained along with a subset of their meta information that includes 
whether the supplied configuration is from a converged calculation as well as the DFT 
code, version, and type of DFT functionals with the total, potential energies. This dataset consists of 
39.1\% Al, 30.7\% Ni, 30.2\% Cu and has 27,987 atomic environments in 3,337 structures.

{\it CHON dataset}: This dataset of molecular structures was extracted from all available 
structures in the NOMAD Archive that only include C, H, O, and N using the NOMAD API. The same procedure 
of selecting only the last entries in each record was applied. This dataset consists of 50.42\% H, 30.41\% C, 
10.36\% N, 8.81\% O and includes 96,804 atomic environments in 5,217 structures.

\section{Analysis Methods} \label{sec:analysis}

\begin{figure*}
\centering
\resizebox{1.0\textwidth}{!}{\includegraphics{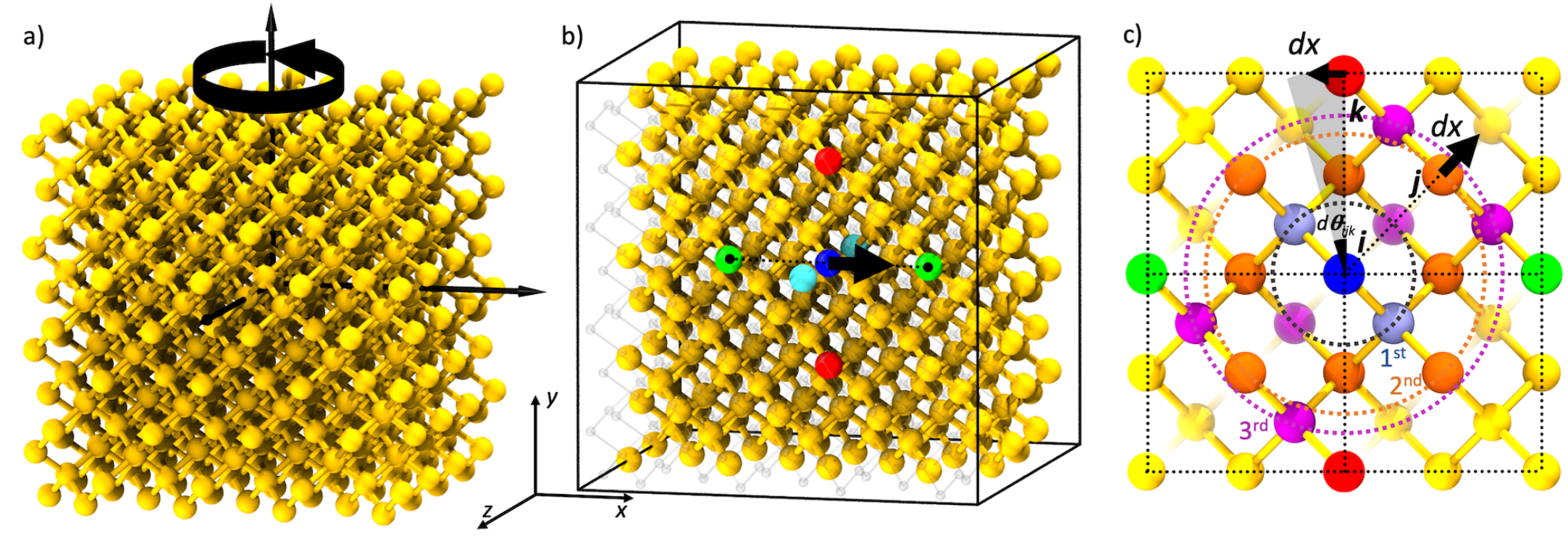}}
\caption{Perturbations on Si structure with $4\times4\times4$ conventional diamond unitcells. (a) Rotation of structure around $y$-axis with $\theta$ angle, (b) perturbation of central atom (dark blue) along $+x$ direction (Mixed Perturbation), and (c) radial and tangential perturbations of atoms at neighbouring shells of central atom. In (c), the cross section of the structure along $x$-$y$ plane is shown with 4 conventional surrounding units cells. Only the atoms within 4 layers along $-z$-axis (up to light blue atom in (b)) can be seen in (c). While green and red atoms are 4$^{th}$-shell neighbours along $x$- and $y$-axes, grey, orange and purple atoms are neighbours at 1$^{st}$, 2$^{nd}$ and 3$^{rd}$ shells, respectively. $d\theta_{ijk}$ shows the angle that corresponds to the tangential perturbation $dx$ on the sphere of 4$^{th}$-shell neighbours.}
\label{fig2}
\end{figure*}

Our analyses for the representations are based on the desired features of encoding 
structural information of the materials that are listed above as {\it invariance}, 
{\it sensitivity}, {\it dimensionality}, {\it differentiability}, 
{\it interpretability}, 
{\it scalability}, {\it complexity}, and {\it discrete mapping}.
While many of these 
features are important according to the application, we focus here on 
the {\it invariance}, {\it sensitivity}, and {\it dimensionality}.

{\it Invariance} of a representation or a descriptor under symmetry operations such as 
translation and rotation is of high concern in developing mapping methods 
since the properties of a material should be identical under these changes of the 
configuration description. The structural representations are constructed to follow 
these conditions otherwise every possible transformation of the material must be included in the training of the machine learning model, leading to unaffordable numbers of permutations. Even a successful construction of such a model can lead to undesired 
predictions for the uncharted permutations that are not in the training datasets. 

{\it Sensitivity} is also an important property of a descriptor since any application needs distinguishable and unique values for the descriptors. 
How sensitive the descriptor is to changes in the structure of the material determines the outcome of similarity analysis or molecular dynamics simulations. For example, if a descriptor produces exactly the same values for any 
perturbation, the outcome will be indistinguishable. 
In MD simulations, such insensitivity will result in inaccurate dynamics due to inaccuracies in energy and force evaluations as a result of artefacts introduced by the descriptor. 

{\it Dimensionality}, on the other hand is more related to the under- or over-determination of the feature space 
for the ML application. While under-determination in the mapping of hyper-dimensional space can 
easily lead to inaccurate predictions of an ML model, over-determination may also lead to undesired predictions 
according to how the over-determined features are eliminated. 

\subsection{Invariance and Sensitivity Analysis}\label{sec:sensitivity}

Our analysis of invariance, local stability and sensitivity are carried out using diamond cubic 
Si structures. All the selected descriptors in this work start from neighbour 
analysis based on atom-centred perspective, as described in 
Section~\ref{section:models} and therefore the descriptors generate values invariant 
under translational symmetries by definition. However, whether they can maintain 
invariance under rotation needs to be verified. To evaluate 
this, we perturbed $4\times4\times4$ cubic diamond crystal Si (c-Si) as follows: rotating the structure around the $y$-axis in a non-periodic unitcell, we calculated the difference of descriptors and 
representations, $dV_\mathrm{all}$ from the 
reference non-rotated structure (see Figure~\ref{fig2}a). 

For the sensitivity analysis, we perform three types of perturbations to a $4\times4\times4$ c-Si unitcell as follows:  

{\bf1. Mixed Perturbation}: the central Si atom with dark blue color in Figure~\ref{fig2}b is moved along the [100] direction (i.e. line joining the light green atoms on $x$-axis) within a periodic supercell by a distance $dx$, ranging from $10^{-8}$\,\AA{} to $0.1$\,\AA{}.

{\bf 2. Radial Perturbations}:  
The atoms in the groups of 1$^{st}$, 2$^{nd}$, 3$^{rd}$, and 4$^{th}$ neighbour atom shells at different distances from the central dark blue 
atom are perturbed along radial and tangential directions (see Figure~\ref{fig2}c). For the radial perturbation, atoms in each shell are moved along the vector that separates the atom from the central atom.
Position change $dx$ ranges from $10^{-8}$\,\AA{} to $0.1$\,\AA{}.

{\bf 3. Tangential Perturbation}:
Neighbouring atoms in the same shells
are perturbed on the sphere inscribed by the distance vector $R_{ij}$ (see Figure~\ref{fig2}c). This perturbation only changes angular contributions of 
the descriptors since the radial distances $R_{ij}$ are kept fixed. Position change ranges from $10^{-5}$\,\AA{} to $0.1$\,\AA{}.

For case 1, we look at the difference in the full structure representation $dV$ (comprising all atomic descriptors) with respected to the unperturbed crystal to investigate whether the full representation is sensitive to small perturbations of a single atom in the structure. In this mixed perturbation test case, all the neighbor distances for the perturbed central atom change from the perspective of the central atom $i$, and the difference $dV$ from the reference unperturbed structure depends on both radial and angular contributions. 

For cases 2 and 3, we look at the difference of the descriptor $dV_i$ of the central atom $i$ with respect to an unperturbed 
neighbourhood. This allows us to test the sensitivity of the individual radial and angular contributions.

\subsubsection{Sensitivity to perturbations}
\label{sec:sensitivitytheory}
We consider the question of whether the representation $V = V(\{R_{ij}\}_j)$ changes in a locally smooth and stable manner near some reference configuration $\hat{R} = \{\hat{R}_{ij}\}_j$. Small changes in the configuration should lead to proportional changes in the representation, a property that we called {\em sensitivity}. This requirement is necessary to represent an arbitrary smooth function with the same symmetries and to inherit its regularity, which in turn is key to obtain accurate fits with few parameters or basis functions.

To illustrate this concept, consider a ``feature map'' $v : \mathbb{R} \to \mathbb{R}$ which is strictly increasing and hence invertible, but assume $v(x) \sim a_2 x^2$ as $x \to 0$. It follows that $x(v) \sim \sqrt{v / a_2}$, i.e. the inverse has a singularity. Suppose now that we wish to represent the linear function $f(x) = x$ as $f(x) = g(v(x))$ then $g(v) = f(x(v)) = x(v)$, i.e., it inherits the singularity which makes it challenging or even impossible to obtain an accurate fit.

In general, we consider paths $R(t) = \{R_{ij}(t)\}_j$ with $R(0) = \hat{R}$ and expand the change in the descriptor to leading order, 
\begin{equation} \label{eq:expand_dV}
    dV = \| V_t - V_0 \| = a_k t^k + O(t^{k+1})
\end{equation}
for some $k \geq 1$. We call a descriptor $V$ linearly stable $\hat{R}$ if $k = 1$ for all possible perturbation paths, i.e., if the change in the descriptor is linear as the perturbation amplitude $t \to 0$. If $k > 1$ then we call $V$ linearly unstable at $\hat{R}$. 

In our sensitivity analyses we choose different perturbation paths $R_{ij}(t)$ leading to different paths in descriptor space $V_t$. From  Equation~\eqref{eq:expand_dV} we then obtain 
\[
    \log \| V_t - V_0 \| \sim k \log t 
    \quad \text{as } t \to 0,
\]
that is we can observe the stability or instability of a descriptor by analysing the slopes on a logarithmic scale. A linear slope, i.e., linear stability is guaranteed to fulfil our {\em sensitivity} requirement. However, in certain high symmetry settings this requirement must be relaxed as we will see in Section~\ref{sec:Sensitivity_Results}.

\subsection{Dimensionality Analysis} \label{sec:dimension_methods}

The descriptors analysed here are all constructed from feature sets extracted from the structural mapping of the atomic neighborhood density to hyper-dimensional spaces. 
Their central objective is the requirement to cover all possible 
perturbations of the structure to ensure a faithful representation to the 
ML model. However, strictly following this concern can lead to 
over-determination in the hyper-dimensional space. In other 
words, the representation may cover only a small subspace of the full    
hyper-space, with the subspace depending on the parameter set used. In 
the case of over-determination, feature sequences may contain many zero entries,  or, for multi-species systems, may need to be padded with zeros to account for species missing from individual environments. Both of these cases lead to high sparsity in the 
descriptors, which in principal could be eliminated by carefully selecting  parameters  
to removing unnecessary features from the final descriptor sets and 
hence from the representations. Moreover, using an
over-determined mapping is likely to induce {\em overfitting} and noise in subsequent ML training. In ML applications, such non-informative 
data should be eliminated before the actual training of the model to reduce 
the error in the training and increase the accuracy of the resulting models.

Due to the well-known {\em curse of the dimensionality}, the dimension of the parameter space of a global optimization problem is the key determiner of the difficulty of obtaining optimal solutions.
When representations form the input data for an optimization problem, their
{\em dimensionality} thus has a crucial role in determining complexity. As the dimensionality increases, the 
number of possibilities rises combinatorially, drastically hindering the task of optimization.

To keep the features at an affordable level for optimization while maintaining an accurate description, one can use 
dimensionality reduction techniques such as CUR decomposition \cite{Imbalzano2018}, farthest point 
sampling (FPS)\cite{Imbalzano2018}, Pearson correlation coefficient (PC) \cite{Imbalzano2018}, or principle component analysis (PCA) \cite{Cubuk2017,Onat2018}. While 
these techniques help to identify the most informative features in the 
descriptors, they also help to analyse how the features of the descriptors 
can provide sufficiently informative data through an analysis of its ``compressibility'', and whether the representation leads to an over-determined embedding. Here, we have used CUR and FPS as implemented in Ref.~\citenum{Imbalzano2018} 
as well as PCA to select the optimum number of features for the descriptors. 

\subsection{Analysis of Dimensionality with Regression} \label{sec:regression_methods}
A key question for models based on 
representations is  how the outcome of predictions are 
depends on the dimensionality of the representation. 
To assess this, we analysed the model 
prediction errors of representations on the structures and total potential energies, $E_\mathrm{DFT}$ taken from the 
 Si GAP fitting database in Ref. \citenum{Bartok2018}. This is a subset of the
complete {\it Si dataset} used elsewhere in this work, chosen here to allow  comparison 
with results for the same dataset in Refs. \citenum{SHIPs}, \citenum{Bartok2018} and \citenum{VanderOord2019}.
For the target values ${\bf t}$ in Equation~\ref{eq1} we use the cohesive energies, i.e.
\begin{equation}\label{eq:cohesive}
t_k = E_\mathrm{DFT}(\chi_k) - N^\mathrm{at}_k E_0, \; k = 1 \dots N_k
\end{equation}
where $N^\mathrm{at}_k$ and $E_0$ are the number of atoms in 
each $\chi_k$ structure and the energy of an isolated Si atom, respectively.
The cohesive energy of structure $\chi_k$ can than be estimated 
via the linear combination
\begin{equation}\label{eq:model}
E_c(\chi_k) = \sum_{v=1}^{N_f}c_{v} V_{k,v}(\chi_k)
\end{equation}
where the index $v$ runs over all
$N_f$ features within the representation.

For a global representation of a structure such as in MBTR the ${\bf V}_k$ 
are taken directly as the representation vectors for each structure $\chi_k$. For 
atom-centered descriptors, however, we first construct a global representation by
summing over the descriptor vectors for each atom ${\bf V}_{k}^{(i)}$, i.e.
\begin{equation}\label{eq:model2}
{\bf V}_k(\chi_k) = \sum_{i=1}^{N^\mathrm{at}_k} {\bf V}_{k}^{(i)} (\chi_k)
\end{equation}
In both cases we normalised the ${\bf V}_k$ representations using the training 
dataset for each representation method.

The cohesive energies can now be estimated by minimizing the 
$L_2$-regularised quadratic loss function
\begin{equation}\label{eq:model4}
J = \sum_{k=1}^{N_k} |E_c(\chi_k) - t_k|^2 + \lambda \sum_{v=1}^{N_f} |c_v|^2
\end{equation}
To estimate the coefficients ${\bf c}$, we build two regression models based 
on (i) linear ridge regression with $L_2$-norm regularisation, 
obtaining the least-squares solution with the QR method (RR)
defined in Refs. \citenum{SHIPs} and \citenum{VanderOord2019} 
and (ii) kernel ridge regression (KRR) as detailed in Ref. \citenum{Jager2018}. 

For the linear ridge regression case,
the regularised least squares problem becomes
\begin{equation}\label{eq:model5}
\min_{{\bf c}} \|{\bf c} \bm\Psi - {\bf t}\|^2_2  + \lambda \|{\bf c}\|^2_2
\end{equation}
where ${\bf t}$ is a vector comprising all the target cohesive energies, 
${\bm \Psi}$ is an $N_k \times N_f$ matrix with representations along each
rows and structures down each column and $\lambda$ is a regularisation parameter.

For the kernel ridge regression case, we 
use the ${\bf V}_k$ representation to build
an $N_k \times N_k$ kernel matrix with elements
\begin{equation}\label{eq:model7}
{\bf K}_{ij} = K({\bf V}_i, {\bf V}_j) = \exp \left(-\gamma \||{\bf V}_i-{\bf V}_j \|^2\right)
\end{equation}
where $\gamma$ is a lengthscale hyperparameter.
After constructing the kernel matrix, one can then predict the
cohesive energy for a new configuration with representation ${\bf V}$ as
\[
    E_c = {\bf c} \kappa({\bf V}),
\]
where ${\bf c} = {\bf t}^T ({\bf K} +\lambda {\bf I})^{-1}$ and $\kappa_k = K(\mathbf{V}_k, \mathbf{V}).$

\section{Results and Discussions}\label{sec:results}

\subsection{Sensitivity}\label{sec:Sensitivity_Results}

\subsubsection{Sensitivity to Rotations}

\begin{figure}
\centering
\resizebox{0.49\textwidth}{!}{\includegraphics{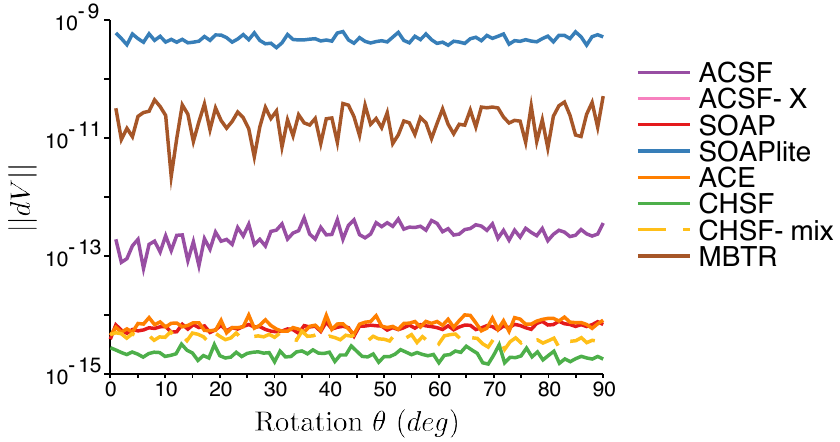}}
\caption{The norm of the difference in representation values between the 
reference structure and rotated whole structure by an angle $\theta$.}
\label{fig3}
\end{figure}

In Figure~\ref{fig3}, we present the norm of the
difference vector $dV$ between the full structure representations 
of the rotated and the reference c-Si system for each approach considered.

Our analyses show that all descriptors maintain the rotational 
invariance with high precision, with all errors below 10$^{-9}$  
(above machine precision $\epsilon$ of $\sim10^{-16}$). Together with built-in invariance with respect to translations and permutation of like atoms of all the representations based on density projections, our analysis indicates 
that all approaches considered in this work fulfill the properties of {\it invariance} in 
rotation, translation and permutation for the structural representations.

Considering the wide range of outcomes on 
the rotation tests, we ascribe the error to 
the numerical precision differences 
i.e. floating point roundoff error of the 
underlying codes. 
For the sake of comparison of the outputs from different approaches
considering the numerical precision of underlying codes, we 
selected $10^{-8}$ as the lower bound in all our subsequent sensitivity 
analysis following the lowest precision observed in these rotation
tests. 

\subsubsection{Sensitivity to Perturbations}
Further analyses are carried out for the sensitivity of 
representations under the atomic motions in structures as described in Section~\ref{sec:sensitivity}. Recall from Section~\ref{sec:sensitivitytheory} that a slope of one indicates local stability (and smooth invertibility of the descriptor) while a slope greater than one leads to singularities in the representation. 

\begin{figure}
\centering
\resizebox{0.48\textwidth}{!}{\includegraphics{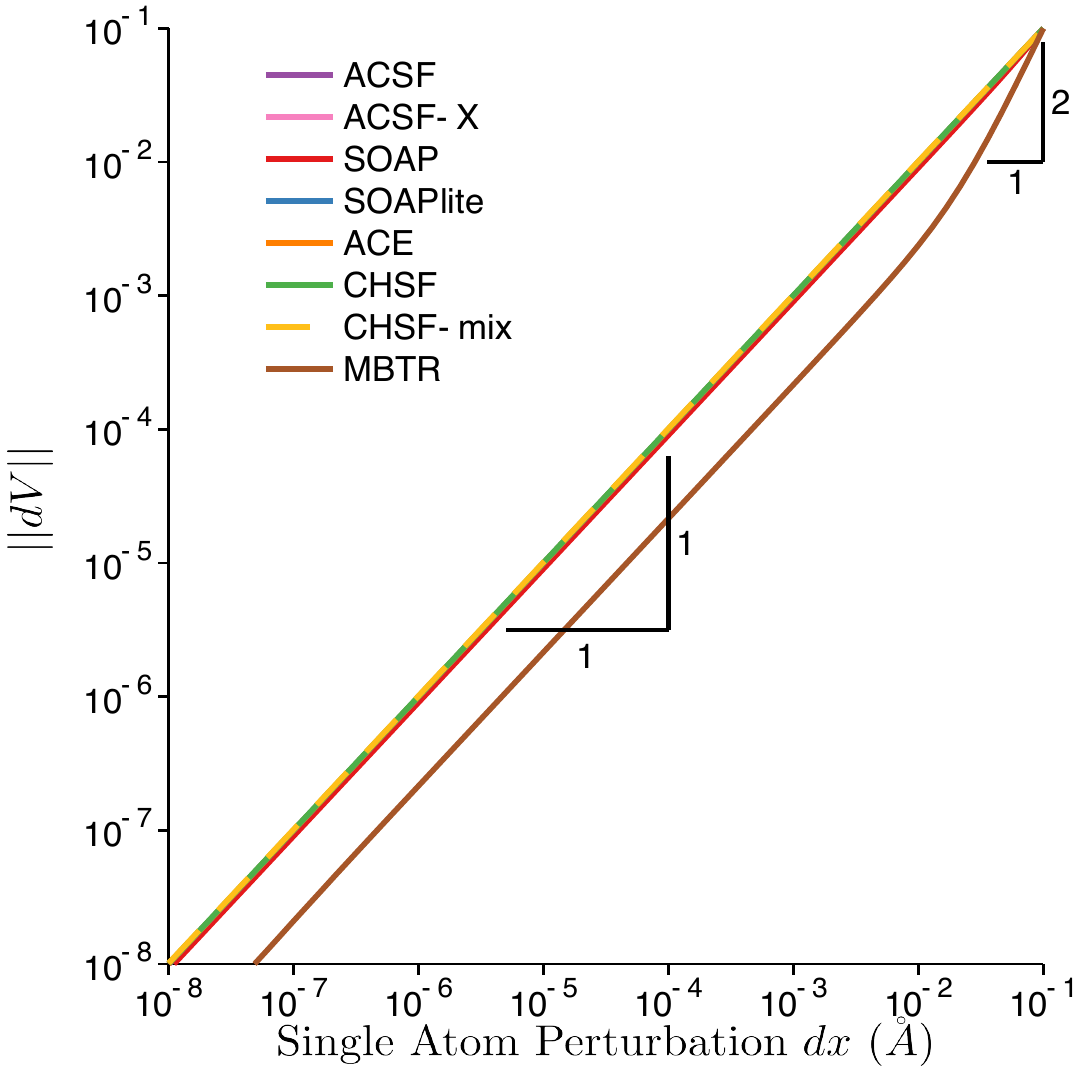}}
\caption{The norm of the difference in representation values between the reference structure and c-Si with perturbation of a single-atom by a distance $dx$, corresponding to the Mixed Perturbation in Section~\ref{sec:sensitivity}.}
\label{fig4}
\end{figure}

Figure~\ref{fig4} shows the results for the Mixed Perturbation with representation changes $\|dV\|$ normalised so that all curves pass through the point (0.1, 0.1) to enable direct comparison.
This lets us to provide an absolute 
comparison for $\|dV\|$ metric between reference and perturbed structures. 
All representations have linear sensitivity within  the entire range of the path, except for MBTR which shows a mild preasymptotic sign of instability (change of slope from 1 to 2 above $0.01$\,\AA{}), which is unlikely to cause any significant deterioration in stability of the representation. 
This is apparent here since MBTR is a 
global representation method. Compared to other atomic descriptors, 
the results for MBTR show that its {\it histogram of angles}
leads to a smooth change in the slope. In the atomic descriptors, this
calculation includes contributions from all descriptors hence the slope is dominated
by radial contributions, which are discussed detailed in next section.

\begin{figure*}
\centering
\resizebox{1.00\textwidth}{!}{\includegraphics{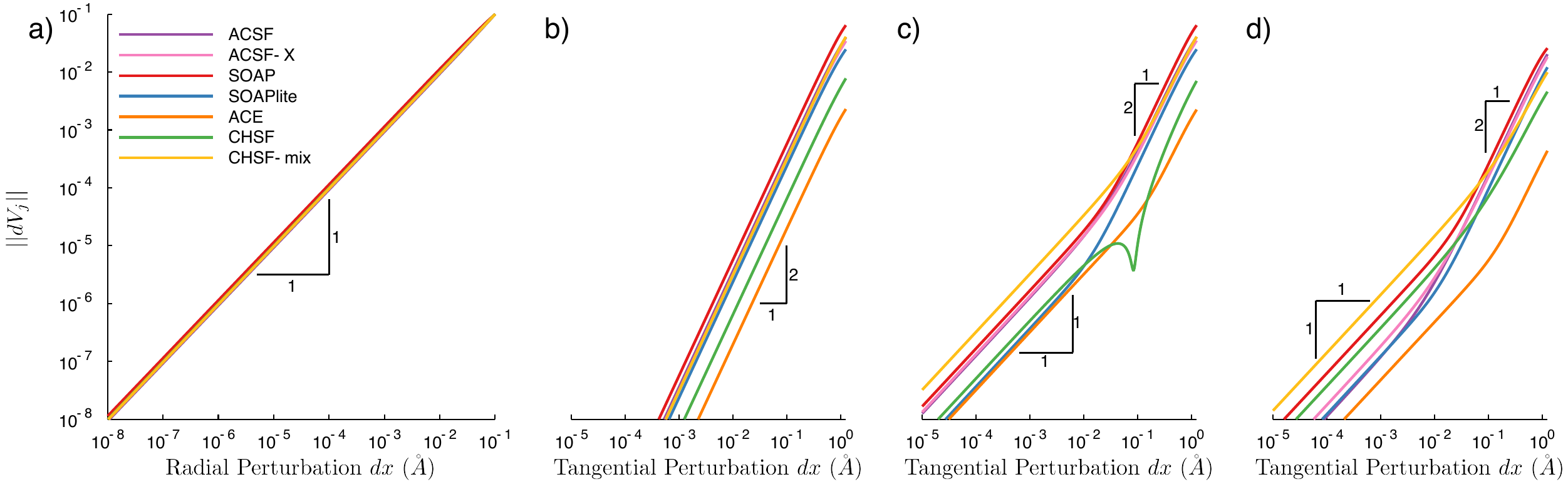}}
\caption{Norm of difference of atomic descriptors on atom $i$ as
a neighbouring atom $j$ is perturbed from its reference position.  
(a) Radial perturbation and (b), (c), and (d) tangential perturbations  
as shown in Figure~\ref{fig2}c.
The tangential perturbations (b) in a high symmetry direction for 1$^{st}$ shell, (c) in a random direction for 1$^{st}$ shell and (d) in a random direction for 2$^{nd}$ shell with random radial perturbations on the same shell before applying tangential perturbation on one of the same shell atom. In (b) and (c), no radial perturbation is applied before the tangential perturbations.}
\label{fig5}
\end{figure*}

\subsubsection{Radial perturbation of reference crystal}
A deeper analysis of the sensitivity of representations 
can be made by analysing the responses of the atomic descriptors to different perturbation modes. 
As described in Section~\ref{sec:sensitivity}, we calculated the change in the descriptor of the central atom $i$ to a radial perturbation of a neighbour $j$ as shown in Figure~\ref{fig2}c. The sensitivity curves corresponding to the radial perturbation of a neighboring atom in the 4$^{th}$ shell are given in Figure~\ref{fig5}a. All descriptors have slope-1 sensitivity curves, indicating linear stability under this perturbation. This is unsurprising since all descriptors provide a relatively high resolution of the 2-body histogram.

\subsubsection{Tangential perturbation of reference crystal} 
Next, we repeat the test of the foregoing section with a tangential perturbation of an atom in the 1$^{st}$ shell. The resulting sensitivity curves are given in Figure~\ref{fig5}b, clearly showing slope 2 for all descriptors.
Thus, according to Section~\ref{sec:sensitivitytheory} {\em all} descriptors are unstable with respect to tangential perturbations, raising concerns due to the resulting singularity in the inverse of the descriptor map. However, the origin of this instability is invariance with respect to reflections about a plane, and fitting any target function with the same reflection symmetry need not be affected by the singularity in the inverse descriptor map.

Concretely, let $i$ denote the centre atom and $k$ the neighbour that is being perturbed in the tangential direction, i.e., 
\[
    R_{ij}^t = R_{ij}^0 + t dR_{ij} + O(t^2),
\]
where $dR_{ij} =  0$ for $j \neq k$ and $\|dR_{ik}\| = 1$ and $dR_{ij} \perp {R}_{ij}^0$. If $R^0$ is symmetric under reflection through the plane that contains the origin and is orthogonal to $dR_{ik}$ (this is the case here) then the configuration $R^{-t}$ is the reflection of $R^t$ to within $O(t^2)$ accuracy. Since all descriptors $V$ we consider are invariant with respect to reflections they necessarily satisfy $\frac{d}{dt}V^t |_{t = 0} = 0$ (this is true for any function of the distances $r_{ij}$ and the cosines $A_{ijk} = \cos \theta_{ijk}$), and hence $V^t \sim a_2 t^2$ as $t \to 0.$ In particular, the inverse descriptor map $V \mapsto R$ must contain a square-root singularity along the path $V^t$.

On the other hand, assume we aim to represent a property, e.g., site potential, $\epsilon = \epsilon(\{R_{ij}\}_j)$, then $\epsilon$ will also satisfy this reflection symmetry, which indicates that the square-root singularity is again removed. 
To illustrate this point further we modify the one-dimensional example from Section~\ref{sec:sensitivitytheory}: 
Assume that we wish to represent $f(x) = x^2 = g(v)$, then $g(v) = f(x(v)) = x(v)^2 \sim v / a_2$ as $v \to 0$, i.e., the singularity is removed in this case. More generally, this occurs whenever $f(x) \sim b_2 x^2$ as $x \to 0$, i.e., when $f$ is symmetric about the origin to leading order.

\subsubsection{Tangential perturbation of a perturbed crystal}
The analysis of the previous paragraph suggests that any perturbation in the configuration of the structure that breaks the symmetry should lead to the linearly stable slope-1 cases. We therefore test which descriptors are capable of capturing this symmetry breaking. 
We break the symmetry in several ways: we perturb an atom in a random tangential direction; we perturb atoms in the second shell which doesn't exhibit the same reflection symmetry; and we perturb the reference crystalline structure in the radial direction from a chosen centre atom $i$, before applying these tangential perturbations.  The results are shown in Figure~\ref{fig5}b--d.

As predicted, any such symmetry breaking leads to changes in the slopes of sensitivity curves of descriptors in the limit $t \to 0$. However, there are differences across descriptors how well the symmetry breaking is captured. First, there are some variations across descriptors how significant the pre-asymptotic slope-2 regimes are, which indicate a reduced sensitivity. However, the most concerning effect is the ``dip'' in the CHSF descriptor in Figure~\ref{fig5}c, highlighting a region of significantly reduced sensitivity (it can almost be thought of as {\em blindspot}) for atomic  displacements in the descriptors, where the perturbation does not change the output values of representations. To test whether adding additional features can remove this dip we implemented an extended CHSF descriptor, labelled CHSF-mix, for which the radial and angular histograms are fully mixed giving a similar description of the 3-body histogram as SOAP and ACE do. This addition clearly removes the reduced sensitivity regions.

\subsection{Dimensionality of Representations}

\begin{figure*}
\centering
\resizebox{1.001\textwidth}{!}{\includegraphics{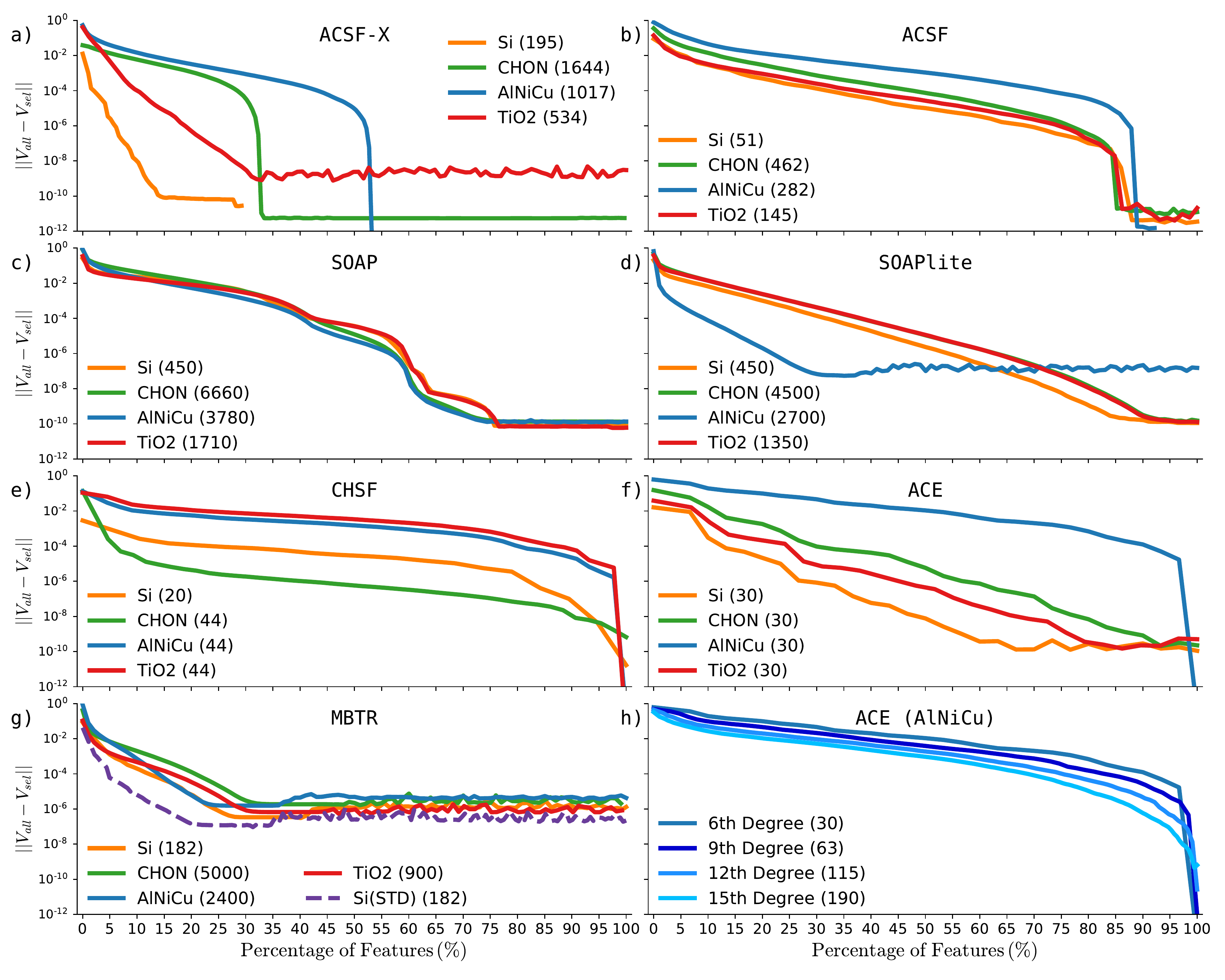}}
\caption{
CUR decomposition based dimensionality reduction analysis of (a) ACSF-X, (b) ACSF, (c) SOAP, (d) SOAPlite, (e) CHSF, 
(f) ACE, and (g) MBTR representations with four datasets that are indicated in legend with the number of features of each within parentheses. Figures show  
the error in the representation from the non-reduced feature set as a function of the percentage of features selected. Legends show the 
corresponding datasets of {\em Si}, {\em CHON}, {\em AlNiCu}, and {\em TiO$_2$} while in panel (g) Si(STD) shows the standardized representation output for {\em Si} dataset and 
in (h) the legend shows the results of ACE with different polynomial degrees from 6 (as used in panel (f)) to 15 for the {\em AlNiCu} dataset.}
\label{fig6}
\end{figure*}

In the second phase of our analyses, we consider four different datasets selected from those described in  Section~\ref{sec:datasets}, namely 
{\em Si}, {\em CHON}, {\em AlNiCu}, and {\em TiO$_{2}$}. Each dataset contains
a diverse range of configurations with thousand of structures. To 
identify how the dimensionality of the representations change with different datasets using the same parameters, we used CUR and FPS feature selection 
techniques and analysed the reduced dimensions of the representations by comparing them with the outcomes of PCA calculations. 
This analysis can also be accounted as a measure of the
\emph{compressibility} of each representation.

\begin{table}[!htb]
\centering
\caption{Number of features for each representation in different datasets. The numbers 
in parentheses show the full feature set before non-zero elements are selected.}
\begin{ruledtabular}
\begin{tabular}{lcccc}
 Desc.     &  Si   &  CHON     & AlNiCu    & TiO$_{2}$ \\ 
\hline
ACSF       & 51      & 462       & 282       & 145   \\
ACSF-X     & 57(195) & 634(1644) & 544(1017) & 534   \\  
SOAP       & 450     & 6660      & 3780      & 1710  \\
SOAPlite   & 450     & 4500      & 2700      & 1350  \\
CHSF       & 20      & 44        & 44        & 44    \\
ACE        & 30      & 30        & 30        & 30    \\
MBTR       & 182     & 5000      & 2400      & 900
\label{tab2}
\end{tabular}
\end{ruledtabular}
\end{table}

\subsubsection{CUR decomposition}

\begin{figure*}
\centering
\resizebox{1.001\textwidth}{!}{\includegraphics{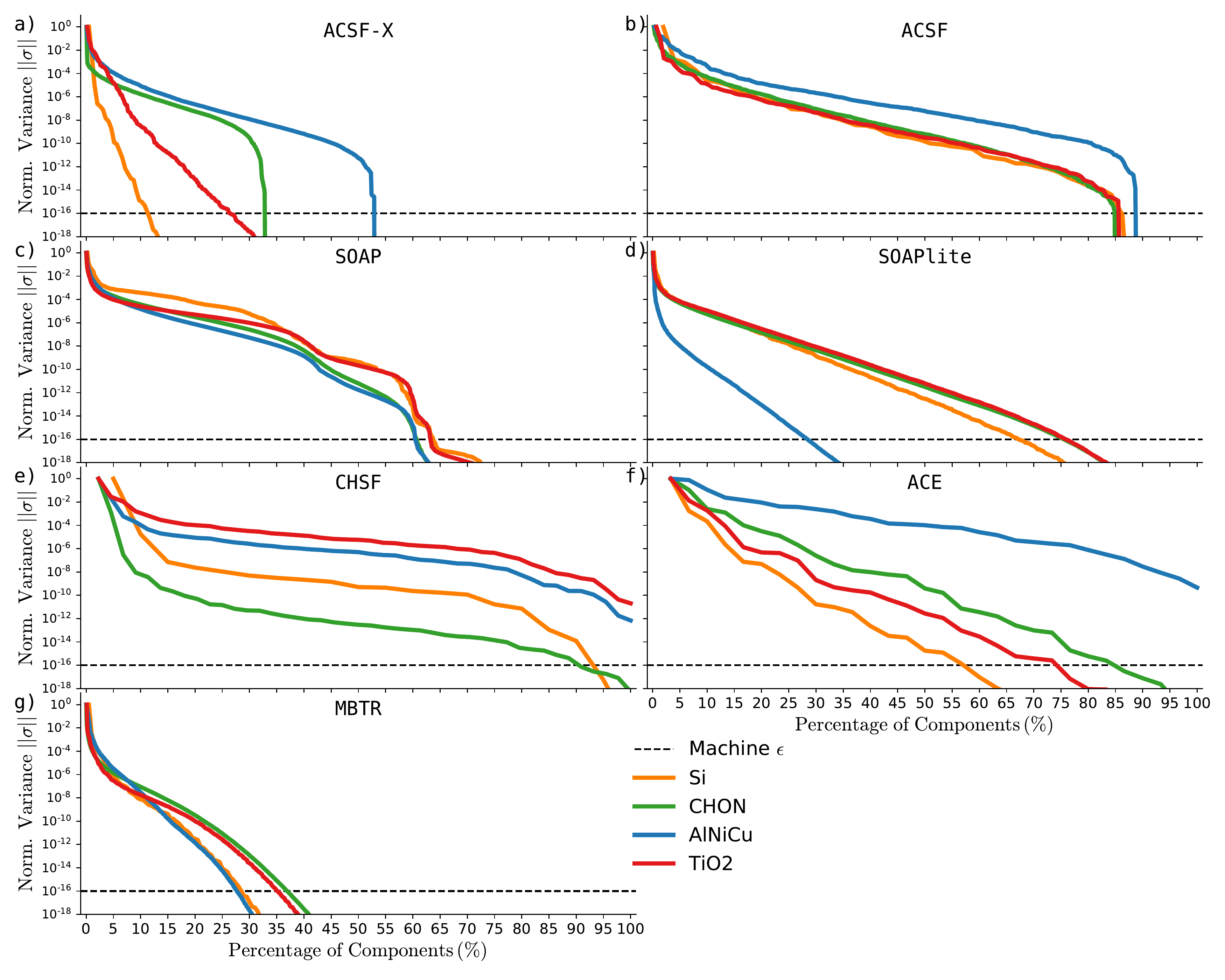}}
\caption{The variance of features versus the percentage of the selected components of (a) ACSF-X, (b) ACSF, 
(c) SOAP, (d) SOAPlite, (e) CHSF, 
(f) ACE, and (g) MBTR representations using 
PCA analysis. Different colors show results 
for each representation with selected datasets. The dashed black and blue lines show the machine
precision and the lower bound in our variance analyses where any value below this threshold is treated as zero. 
}
\label{fig7}
\end{figure*}

As a first step we analysed the representations including all element-wise descriptors. MBTR is considered as 
a representation-only approach as its output cannot be broken down to element-wise descriptors.  In Figure~\ref{fig6}, the total error between the full feature sets of each dataset 
and the reduced feature sets that are extracted from CUR analysis is presented. As 
each approach provides a different number of features for the datasets at hand depending on the 
selection of (hyper)parameters, we provide a complete list of the number of 
non-zero features in each representation with the corresponding datasets in Table \ref{tab2} and in the legends of each panel of Figure~\ref{fig6}. 

After removing any features of ACSF-X that are all zero from {\em Si} and {\em AlNiCu} dataset (see Figure~\ref{fig6}a), the selection method CUR in 
Section~\ref{sec:dimension_methods} is applied throughout the dataset and features from the full feature set are 
selected one-by-one and added to the new feature set by
calculating the error with respect to the full feature 
representation. Using this method, the 
features that contribute most to the representation can be selected. As the lower contributions are added to the new feature set, the error cannot 
be reduced more and the overall error becomes constant, equal to zero within numerical precision. The number of 
features that are selected at the beginning of this plateau can be counted to determine the size of the compressed representation, and conversely the number of remaining features can be thought of as the over-determination of the representation.

In Figure~\ref{fig6}, one can see three types of results: 
(i) those with error curves that gradually decrease to the point where the error plateaus as for the {\em TiO$_2$} results of ACE, ACSF-X and MBTR; 
(ii) errors decrease step-wise to a plateau such as in SOAP results; and 
(iii) where errors drop rapidly
as for the {\em AlNiCu} results in ACSF, ACSF-X, ACE and CHSF. 

SOAP and SOAPlite, in Figure~\ref{fig6}c and d, respectively, can be directly compared since they differ only in the choice of radial basis function. The results for the {\em Si}, {\em CHON}, and 
{\em TiO$_2$} datasets can be examined for both approaches. While SOAPlite has close to exponential decay up to $\sim$70\% of selected features, in SOAP this regime extends only up to 30\%. After this, the SOAP error has a more step-like character. 
Similar behaviour can also be seen in type (iii) results such as
ACSF, ACSF-X, CHSF, and ACE. For these methods, while the first $\sim$10\% of features vary the error reduction, the rest of the features only slightly reduce the error. 
For the {\em AlNiCu} dataset, SOAP and SOAPlite have substantially different outcome from the rest of the datasets which is an 
indication on how the choice of radial basis can be a key 
determiner for the relevance of features in the representations.
One can also see the dominance of radial basis in the global 
representations when Figure~\ref{fig4} is compared with Figure~\ref{fig5}. The optimal choice of basis functions
for descriptor performance is another open question and is outside the scope of present work. We 
leave further investigation of basis functions' role on the sparsification and descriptor performance for future work.

\begin{figure*}
\centering
\resizebox{1.001\textwidth}{!}{\includegraphics{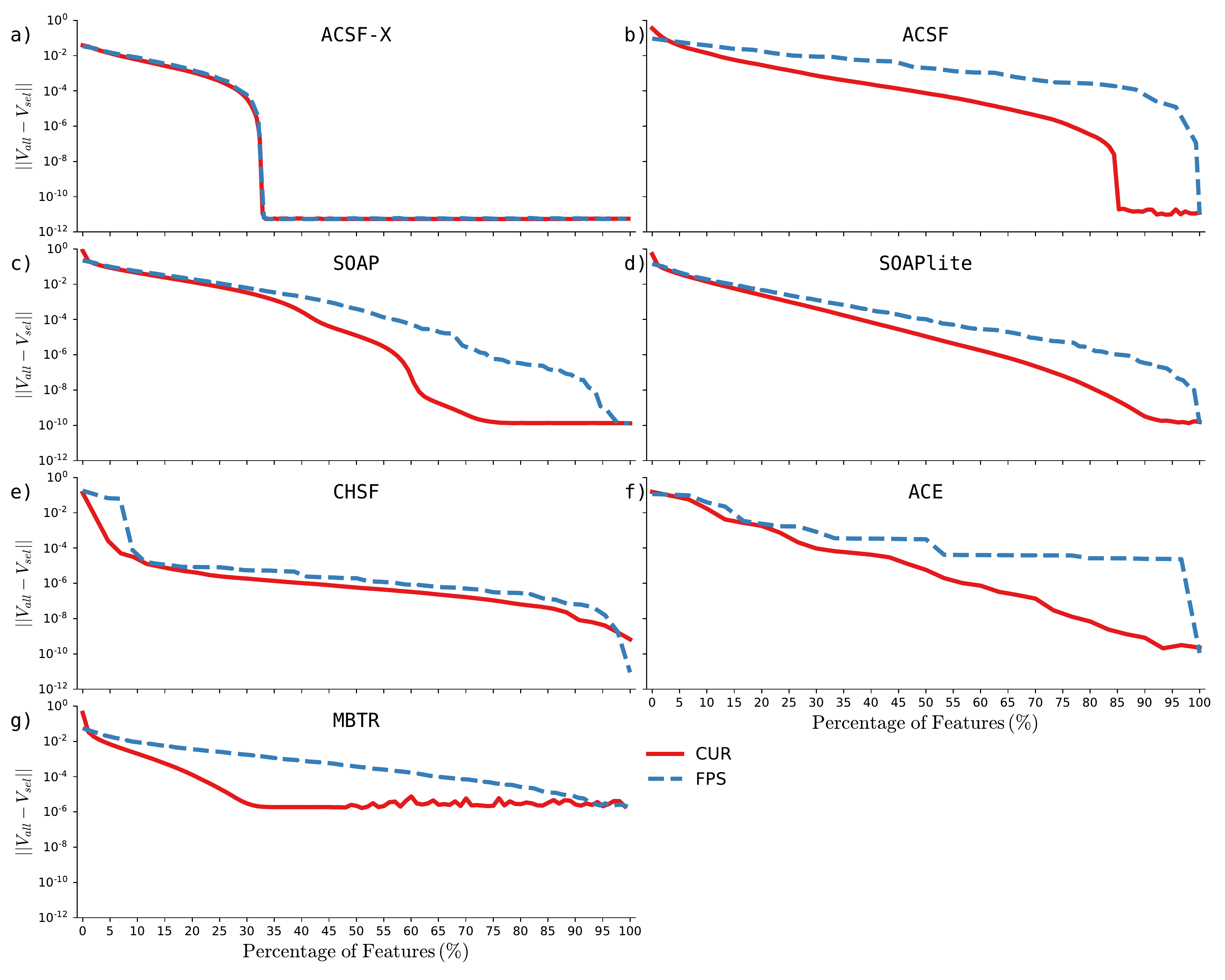}}
\caption{Comparison of CUR and FPS based dimension reduction analysis on ACSF both with 
(a) extended and (b) standard parameter sets, (c) SOAP, (d) SOAPlite, (e) CHSF, 
(f) ACE, and (g) MBTR representations for {\em CHON} dataset.}
\label{fig8}
\end{figure*}

In most of our dimensionality reduction results, one can see distinct constant error regions associated with over-determination. These features should ideally be eliminated 
before using the representation in ML models. For example, consider the results for ACSF-X 
and ACSF in Figure~\ref{fig6}~a and b. As the aim of ACSF-X is to 
extend the number of features in set from the widely-used and well-tested standard ACSF parameter set, some of 
the features are 
expected to be irrelevant for representing structures in our datasets. It is thus unsurprising that the dimension reduction analyses in Figure~\ref{fig6}a show that there only fractions of 
features contribute significantly - around 14\% for {\em Si}, 33\% for {\em CHON}, 54\% for {\em AlNiCu}, and 33\% for 
{\em TiO$_2$}. 
A smoother feature reduction can be seen for the standard ACSF parameter set in Figure~\ref{fig6}b, where the error decay is very similar with about 85\% of the 
total features sufficient across all four datasets and thus around 15\% of redundant features.

A similar result is seen for SOAP, where around 75\% of the features of SOAP representations are sufficient to cover the structural 
variance across all datasets. This result is more striking than that for ACSF since SOAP has about an order 
of magnitude more features in its representation. We can conclude that 
both ACSF and SOAP are robust approaches that cover the hyper-dimensional space of 
structural representation for a wide-range of crystals and molecules.

The ACE {\em AlNiCu} results are significantly different than the other datasets. 
To identify whether the degree of the polynomial is the reason for this pattern, we carried 
out additional analyses with ACE, increasing the degree of polynomial from 6th to 15th 
degree in steps of 3 degrees, as shown in Figure~\ref{fig6}h. 
Increasing polynomial degree significantly increases the number of features; however, we find that the percentage of selected features on the final representation does not change significantly 
and the pattern of error decay is still quite different from the rest of the datasets (e.g, {\em Si}, where 25\% of the features can be removed from the representation although it has order of 
magnitude less features in the descriptor vectors than with 15th degree of polynomial expansion.

To further investigate the extensive redundancy identified for MBTR features across all four datasets, we consider whether the discretised smearing of positions and angles used by the MBTR representation leads to clustering of 
the features representation space. To identify any 
clustering of the data, we perform standardization of the features for all the 
representations of {\em Si} that are generated by MBTR and show the feature 
selection curve in Figure~\ref{fig6}g labelled as Si(STD). When compared with {\em Si} curve, Si(STD) 
does reduce the error when less than $\sim$20\% of features are selected by around an order of magnitude in comparison to the raw representation. However, 
standardization suggests an even smaller feature set selection of about only 20\% of the full feature set. This may be due to the Gaussian smearing with $D$ in MBTR that significantly increases the correlation between features. 

\begin{figure*}
\centering
\resizebox{1.001\textwidth}{!}{\includegraphics{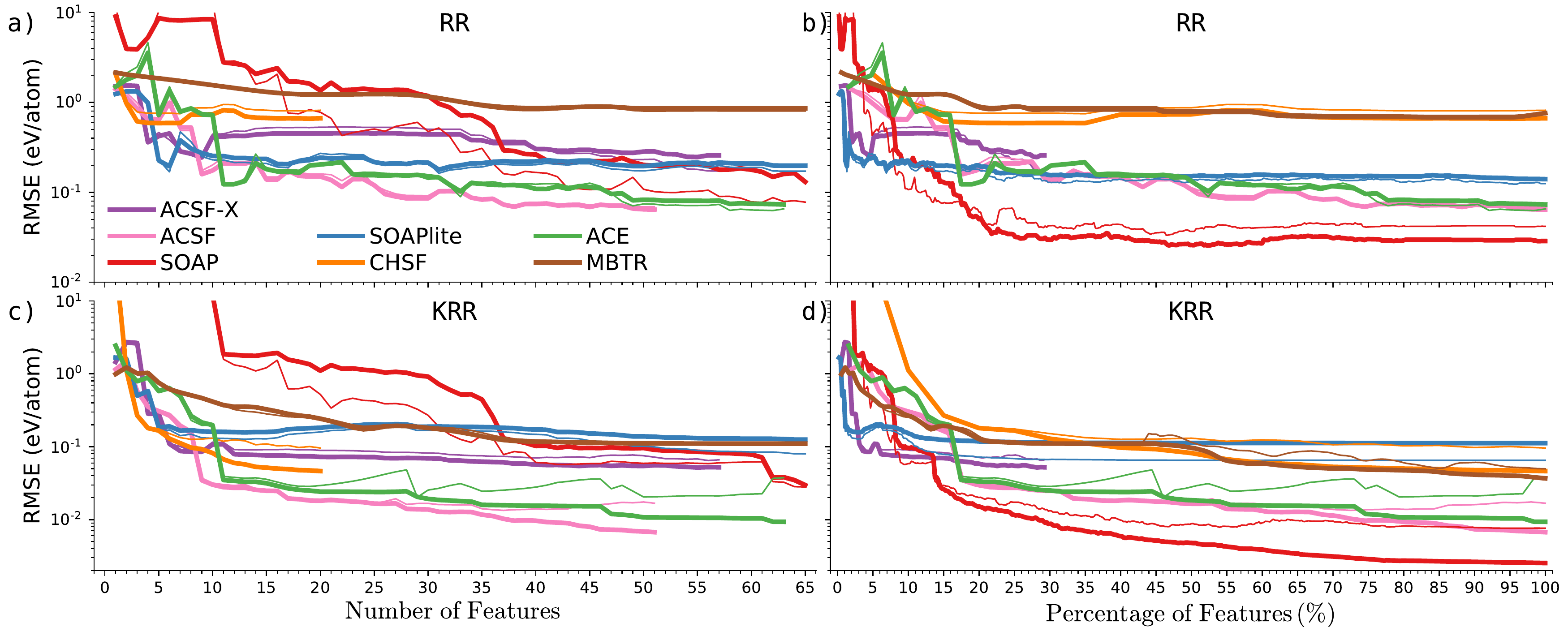}}
\caption{Comparison of RMSE using (a),(b) RR and (c),(d) KRR with Si dataset on representations while (a) and (c) showing up to the top 65 features that are selected with CUR method. Thick lines represent training errors, thin lines represent test errors.}
\label{fig9}
\end{figure*}

\subsubsection{Principal Component Analysis}

The CUR selection process is closely related to the 
{\em principle components} of the representation data for each dataset. To show 
whether these principle components are related to the final feature selections in CUR, we further analysed
the datasets using PCA. In Figure~\ref{fig7}, the fraction of variance explained by the 
principle components of the four datasets are given for each representation. 
Since the PCA variances decrease from one to very small values, we determine a 
lower bound after which we consider the variance to be zero, shown as the
{\em zero level} in our figures. PCA variation results for principle components  
follow the same trends as the CUR curves discussed above. Although CUR results show directly the 
hyper-dimensional space of the representation, PCA results are not solely 
based on the selected features but are a collective property of all features based on the covariance matrix from which the principle components are extracted. 
As seen in Figure~\ref{fig7}, 
the outcome of this selection following the highest to lowest variances and 
extraction of the features with highest values in the covariance matrix 
gives similar results to the CUR decomposition.

\subsubsection{Farthest Point Sampling}

CUR and PCA are both based on SVD decomposition, selecting the features 
as orthogonal dimensions of the hyper-dimensional space, treating each feature 
as linearly independent since linearly dependent features cannot span the space.  However, neither of
these methods consider if the selected features have non-linear dependencies 
on other features. Another approach to select features without using SVD decomposition is Farthest Point Sampling (FPS).  In FPS, the feature selected at each iteration is chosen as the farthest from those already selected.
Hence, FPS does not provide information on whether the selected feature 
is indeed linearly independent to those already selected. This can be understood by considering each features' distance from the others at a time late in the process when there are few remaining features to be selected. These remaining features either represent very small distances 
as minor additions to the previously selected and clustered features or repeat 
similar distances.

In Figure~\ref{fig8}, the FPS results for the feature selection show 
that there are significant differences in error reduction and hence dimension 
compression for SOAP, SOAPlite, ACE and MBTR representations in comparison to our earlier CUR results. However, these FPS results 
do not allow insight into each features' contribution as a dimension in the hyper-dimensional 
space. The relatively small reduction of errors in FPS selection or the constant regions are due to its selection criteria, which is based only on hyper-spatial distance. This poses a limitation 
in the analysis if one would like to find the full extent of the representation and 
remove non-informative features from the descriptors. However, determining the 
cutoff for the features according to the error is not obvious since there may not be 
a plateau in the error --- as is seen for in ACSF-X, where only 25\% of features contribute --- but instead a more gradual decrease as in the results for all other representations.

\subsection{Regression with Representations}

The RR and KRR methods are applied to all representations using 
subset of {\it Si dataset} (see Section~\ref{sec:regression_methods}) and 
RMSE is calculated as a function of the number of features in each representation.

For each representation, the regularisation parameter 
$\lambda$ used in both RR and KRR methods and the $\gamma$ 
hyperparameter of ${\bf K}$ in KRR were chosen
to optimize the convergence of the RMSE in the predictions on an independent test set.

In Figure~\ref{fig9}, we present our regression results for the two 
ridge regression methods RR and KRR in the first and second rows 
of the figure, respectively. While RMSE of both training (thick lines) 
and test datasets (thin lines) are shown in Figure~\ref{fig9}a and c 
for each regression method and for the top 65 features that selected by the CUR 
method, the RMSE of full feature extension of representations 
are provided for comparison in Figure~\ref{fig9}b and d.

Our intention here is not to provide the best potential energy 
surface (PES) estimator for each representation method, since this has already 
been extensively studied in the literature but instead to provide a 
common metric to compare different representations.
Here by fitting those methods to same dataset, we can assess how each method 
performs under this task, and in particular how the dimensionality of each method 
effects the outcomes.

The effect of the dimensionality of the feature space can be seen from 
the RR outcomes of representation for the full feature set. 
In Figure~\ref{fig9}b, one can identify plateaus at three typical
RMSE values: 0.9, 0.1, and 0.04 eV/atom. While RMSE for 
both training and test dataset of SOAP has the highest accuracy with 0.04 
eV/atom, MBTR and CHSF have the lowest accuracy of around 0.9 eV/atom. The rest 
of the methods have final errors of around 0.1 eV/atom. As the ridge regression is a linear 
fit to representation features, the RMSE can be 
compared directly between methods. The results indicate that SOAP has enough features to cover the structural 
variation but convergence is slow using a linear fit. 
However, using a non-linear fit as in Figure~\ref{fig9}d outperforms
RR. While SOAP has still the highest accuracy, albeit with a 
significant split of RMSE between test and training datasets, ACSF, ACE, 
and SOAPlite result in predictions that are at least an order of magnitude higher in accuracy in KRR than in RR.

To identify the role of the most important features in each representation on the prediction RMSE, 
we analysed the RMSE of methods while only using the top-ranked selected features from 
CUR up to 65 features. The RMSE of RR shows that ACE and ACSF have
very similar results up to 50 features, where they reach an accuracy comparable to that of 
65 features with the SOAP or SOAPlite models. 
When non-linearity is introduced in the KRR models in Figure~\ref{fig9}c the difference is even more stark, for example CHSF with 
only 20 features has comparable accuracy with SOAP at 65 features. 

These results raise the question: what is the best feature set 
for a given representation, or equivalently, what is the necessary 
dimensionality for a given representation for a specific 
dataset or material? 
In Figure~\ref{fig9}d, we can see that RMSE of training and test datasets for 
SOAP and ACSF start to diverge at about 50\%, and 70\% of features, respectively. These  
points coincide with the region in Figure~\ref{fig7} where the variance of these
representations drops below about $10^{-12}$. Similar results can also be observed from the KRR results 
for MBTR, CHSF, and ACE where either RMSE does not reduce more with the 
addition of new features such as in ACE beyond 45\% 
features, or RMSE of test dataset significantly increases 
as in MBTR after 40\% features are selected.

These results show that there are clear inconsistencies between the 
selections of features in representation and even within the same 
representation as two critical points need to be considered 
before applying a feature set to MLIPs using dimension reduction: 
1) the number of features may not be indicative of complete
coverage of structural representation of a method, 2) increasing 
the number of features may not result in gaining benefit and may 
also cause overfitting.

\section{Conclusion}

We have carried out a comprehensive
assessment of the sensitivity 
of atomic environment representations, using several methods to analyse the sensitivity under rotation and various perturbations.
Our results show that although many representations provide an overall acceptable  
accuracy for sensitivity, there is still room to balance sensitivities to radial and angular perturbations.
We thus conclude that further investigation of how insensitivities effect applications of interatomic potentials and hence observables in MD simulations is necessary to improve 
ML driven simulation approaches.

We also carried out extensive dimensionality analyses of various representations, which have identified significant opportunities to eliminate unnecessary information that may reduce the accuracy of predictions from ML models. 
We also conducted regression tests to provide a comparison between representations as their dimensionality varies. The results show clear differences in the number and fraction of important dimensions in the different representations.
This is expected to become increasingly important as more complex representations are developed, and especially when incorporating property-based descriptors alongside atomic environment representations.

\section{Data Availability Statement}

The data that supports the findings of this study are openly available in
github.com/DescriptorZoo/Materials-Datasets at http://doi.org/10.5281/zenodo.3871650, Version v1.0 that are extracted from open-access NOMAD Archive (http://nomad-coe.eu)\cite{DATASHARE}. The details of all datasets are given at Section~\ref{sec:datasets}. The corresponding citations for other data that are used in this study are available from the following publications: GAP Si potential database from Ref. \cite{Bartok2018}, Si molecular dynamics (MD) database from Ref. \cite{Cubuk2017} and  TiO$_2$ dataset from Ref. \cite{Artrith2016} through web access from Ref. \cite{AENETdata}. The codes that are used to create representations are detailed at Section~\ref{section:codes}.

\begin{acknowledgments}

We thank G\'abor Cs\'anyi and Albert Bartok-Partay for useful discussions. This work was supported by grants from the Leverhulme Trust under Grant RPG-2017-191. This work received funding from the EU’s Horizon 2020 Research and Innovation Programme, Grant Agreement
No. 676580, the NOMAD Laboratory CoE.

We used data from the Novel Materials Discovery (NOMAD) Laboratory (\url{https://nomad-coe.eu/}).
We are grateful for computational support from the UK national high performance computing service, ARCHER, for which access was obtained via the UKCP consortium and funded by EPSRC grant reference
EP/P022065/1. Additional computing facilities were provided
by the Scientific Computing Research Technology Platform of
the University of Warwick.

\end{acknowledgments}
 
\bibliography{DescSensDim-v2}

\end{document}